\documentclass[journal,onecolumn,11pt]{IEEEtran}

\usepackage{cite,epsfig,times,amsmath,amssymb,psfrag,graphicx}

\newtheorem{theorem}{Theorem}

\newtheorem{lemma}{Lemma}

\newtheorem{numex}{Example}
\newtheorem{remarks}{Remark}
\newtheorem{definition}{Definition}

\def\s{\mathbf{s}}
\def\S{\mathbf{S}}
\def\r{\mathbf{r}}
\def\R{\mathbf{R}}
\def\v{\mathbf{v}}
\def\x{\mathbf{x}}
\def\X{\mathbf{X}}
\def\d{\mathbf{d}}

\def\e{\mathbf{e}}
\def\n{\mathbf{n}}
\def\t{\mathbf{t}}
\def\dw{\widetilde{d}}
\def\dwb{\widetilde{\mathbf{d}}}
\def\xw{\widetilde{x}}
\def\sw{\widetilde{s}}
\def\Xw{\widetilde{X}}
\def\Sw{\widetilde{S}}
\def\xwb{\widetilde{\mathbf{x}}}
\def\Xwb{\widetilde{\mathbf{X}}}
\def\Swb{\widetilde{\mathbf{S}}}
\def\Xw{\widetilde{X}}
\def\zero{\mathbf{0}}
\def\one{\mathbf{1}}

\newcommand{\lambdab}{\mbox{\boldmath$\lambda$}}

\newcommand{\xib}{\mbox{\boldmath$\xi$}}

\newcommand{\eq}{\begin{equation}}
\newcommand{\en}{\end{equation}}

\centerfigcaptionstrue

\begin{document}

\title{Packet Scheduling in Switches \\
with Target Outflow Profiles}

\author{
\authorblockN{Aditya Dua} \\
\authorblockA{Qualcomm Inc.}\\
\authorblockA{3165 Kifer Rd., Santa Clara, CA 95051}\\
\authorblockA{{\it adua@qualcomm.com}}\\
\authorblockN{Nicholas Bambos}\\
\authorblockA{Dept. of Elec. Engg. and Dept. of Mgmt. Sci. and Engg.}\\
\authorblockA{Stanford University}\\
\authorblockA{350 Serra Mall, Stanford, CA 94305}\\
\authorblockA{{\it bambos@stanford.edu}}\\
}

\maketitle

\begin{abstract}
The problem of packet scheduling for traffic streams with target outflow profiles traversing input queued switches is formulated in this paper. Target outflow profiles specify the desirable inter-departure times of packets leaving the switch from each traffic stream. The goal of the switch scheduler is to dynamically select service  configurations of the switch, so that actual outflow streams (``pulled" through the switch) adhere to their desired target profiles as accurately as possible.

Dynamic service controls (schedules) are developed to minimize deviation of actual outflow streams from their targets and suppress stream ``distortion''. Using appropriately selected subsets of service configurations of the switch, efficient schedules are designed, which deliver high performance at relatively low complexity. Some of these schedules are provably shown to achieve 100\% pull-throughput.  Moreover, simulations demonstrate that for even substantial contention of streams through the switch, due to stringent/intense target outflow profiles, the proposed schedules achieve closely their  target profiles and suppress stream distortion.

The switch model investigated here deviates from the classical switching paradigm. In the latter, the goal of packet scheduling is primarily to ``push'' as much traffic load through the switch as possible, while controlling delay to traverse the switch and keeping congestion/backlogs from exploding. In the model presented here, however, the goal of packet scheduling is to ``pull'' traffic streams through the switch, maintaining desirable (target) outflow profiles.
\end{abstract}



\section{Introduction}
\label{sec:introduction}

Real-time services such as multimedia streaming, video on demand, video telephony etc. continue to gain popularity amongst Internet users. These applications have strict quality-of-service (QoS) requirements with regard to packet delivery times and jitter. Scheduling algorithms employed in packet switches/routers play a key role in QoS provisioning for real-time Internet applications.

While early research on packet switching focused on the output-queued (OQ) switch architecture \cite {parekh1,parekh2}, input-queued (IQ) switches have received much attention in recent times, owing to their scalable architecture. However, non-trivial scheduling/arbitration algorithms are needed to resolve contention between input traffic streams to ensure efficient operation of an IQ switch. Most research on IQ switch scheduling has revolved around performance metrics like throughput and average delay, which are conceived on {\it macro} time-scales (at the mean flow level). Numerous scheduling algorithms based on maximum weight matching (MWM), projective cone schedules (PCS), etc. have been proposed in the literature \cite{nick1}-\cite{ross1}, all of which provably guarantee $100\%$ (push)-throughput, with varying degrees of average delay performance. This body of literature, while important in its own right, does not address the problem of QoS provisioning for time/jitter sensitive real-time traffic, which entails performance engineering and control of the switch on {\it micro} time-scales (at the packet level).

In an initial effort to address the latter problem, in this paper, we develop IQ switch scheduling algorithms for traffic streams associated with {\it target outflow profiles}. The target profile of a  traffic stream specifies the desirable (hence, the term ``target'') packet {\it inter-departure times} (IDT) of packets leaving the switch. In other words, the target outflow profile determines the {\it ideal} packet inter-departure times.

In the absence of congestion, packets from each stream will depart the switch in accordance with the associated target profile. However, contention between competing traffic streams for the shared switch fabric causes congestion in the switch. Consequently, the actual departure process of a stream {\it deviates} from the ideal departure process (as dictated by its target outflow profile). In other words, the stream outflow gets {\it distorted} by the switch, vis-\`a-vis its target profile. Thus, the objective of the switch service scheduler is to minimize the aggregate distortion of the target output profiles of all streams traversing the switch. That is, the scheduler must select switch {\it service traces} (sequences of switch configurations) such that the actual departure/outflow profiles of streams track their corresponding target profiles as accurately as possible. We call this the {\bf Service Trace Control} (STC) problem for an IQ switch.

The motivation behind seeking a solution to the STC problem is to render packet switched networks somewhat ``transparent'' to timing/jitter sensitive multimedia traffic. The target outflow profiles are determined by the times at which consecutive packets need to be delivered to end users to ensure uninterrupted multimedia playout (the playout profile). High quality multimedia experience is provided to end-users if traffic streams negotiate routers/switches with minimal distortion. Note that the term ``distortion'' is simply used in this paper in connection to deviation of packet inter-departure times from their target profiles. The term is {\it not} used as in information theory and coding theory, where it has a deeper meaning.
(deviation from target profiles).

In our switch model, delayed packets are not dropped, but instead are penalized for violating their target packet inter-departure times (IDT). The switch is also penalized for being ahead of the target packet IDTs. This is done to prevent buffer overflows at downstream nodes (flow control) and the end-user, as well to avoid starvation of best-effort traffic (i.e. without target outflow profiles) being served by the switch. This model is representative of half-duplex applications like lossless multimedia streaming (e.g. an online baseball game), where the end-user would much rather wait for a delayed packet than miss viewing the media content encoded in the delayed packet (which would happen if the switch drops delayed packets).

In our framework, packets can be thought of as being associated with {\it soft deadlines} for their inter-departure times (IDT). Any positive deviation (exceeding the deadline) from the target IDTs manifests itself as a soft deadline violation, which carries a penalty/cost. The ``softness'' of a deadline is reflected by the cost associated with its violation (the lower the violation cost, the softer the deadline). On the other hand, any negative deviation from the target IDTs (transmitting before a target inter-departure time) is also a soft deadline violation, and carries a cost (e.g. for stressing downstream receivers with potential buffer overflows). The service trace control (STC) problem thus translates to minimization of aggregate soft deadline violation cost over all traffic streams. 
This is explained in detail in Section \ref{sec:stc}.

In the classical packet switching paradigm (see \cite{nick1}-\cite{ross1})) incoming traffic flows compete for switch service. The scheduler's objective is to control the congestion buildup (and avoid excessive backlogs), given the traffic load. Alternatively, the scheduler tries to maximize the inflow load that can be ``pushed'' through the switch, without the packet backlogs exploding. Hence, it tries to maximize the ``push-throughput''. In the switch model studied in this paper, the issue is very different. Packets streams are ``pulled'' through and out of the switch. The packets initially reside in input queues, organized as virtual output queues (VOQ). Recall that the scheduler's objective is now to pull the streams through and out of the switch, so that their outflow packet inter-departure times (IDT) deviate as little as possible from specified targets and the outflow stream distortion is minimized. But if the target IDTs are too short (outflow target profiles have high intensity) the switch may not be able to keep up and the distortion of one or more streams may grow excessively over time. Thus, the scheduler can now be viewed as trying to maximize the ``pull-throughput'' of the switch, i.e., supply the most intense outflow streams, while keeping their distortions under control. This is explained in detail in Sections \ref{sec:stc} and \ref{sec:admit}.

\subsection{Related work}
\label{ssec:relwork}

The case of scheduling periodic messages through IQ switches has been addressed in the literature.  In that case, packets for each traffic stream are generated periodically, and the maximum time allowed for transmission of a packet is equal to the period of the stream. A schedule is deemed feasible if all messages meet their deadline requirements. Note that the periodic model is a special case of our general model, with constant inter-departure times (equal to the period of the stream). Inukai \cite {inukai} showed that a feasible schedule can be constructed when the periods of all streams are equal and both input and output link utilization are less than 1. Liu et al. \cite {liu} conjectured that Inukai's conclusion holds for traffic streams with arbitrary periods and also proposed heuristic scheduling algorithms based on the earliest deadline first (EDF) and minimum laxity first (MLF) policies. The performance of their heuristics degrades rapidly with switch size. In support of the conjecture, Giles et al. \cite {hajek} proposed the nested periodic scheduling (NPS) rule, which finds a feasible schedule when each period divides all longer periods and link utilization is less than 1. NPS also finds a feasible schedule for arbitrary message periods, provided the link utilization is no more than 1/4. The computational complexity of NPS is ${\cal O}(N^4)$ for an $N \times N$ switch. Rai et al. \cite {rai} developed heuristic weighted round robin (WRR) scheduling policies for multiclass periodic traffic, with an online implementation complexity of ${\cal O}(N^3)$. More recently, Lee et al. \cite{lee} proposed the Flowbased Iterative Packet Scheduling (FIPS) algorithm for periodic traffic with two classes, which minimizes the number of dropped packets when the switch is overloaded. They extended the FIPS algorithm to design efficient heuristics for arbitrary multiclass traffic. Their proposed algorithms outperform MLF and EDF based policies, but have a complexity of ${\cal O}(N^{4.5})$.

On a different strand of research, Li et al. \cite {li} developed a frame-based scheduler with guaranteed delay and jitter bounds for leaky-bucket constrained traffic. Chang et al. proposed schemes for providing delay guarantees in IQ switches based on the Birkhoff-von Neumann (BV) decomposition of the input rate matrix in \cite {chang1} and based on EDF for load balanced switches (see \cite {chang2}) in \cite {chang3}. Their schemes have an offline computational complexity of ${\cal O}(N^{4.5})$ and an online memory requirement of ${\cal O}(N^3 \log N)$. Keslassy et al. \cite {keslassy} proposed a frame based scheduler based on the BV decomposition to guarantee low jitter, under the assumption that jitter sensitive traffic forms a small fraction of the overall switch load.

A common feature of the above works is that they deal with scheduling of smooth/regular traffic (completely characterized by a single fixed rate known to the scheduler). However, traffic arriving to a switch can be irregular due to the bursty nature of traffic sources (e.g. variable bit rate video), due to flow aggregation, or due to jitter induced by upstream switches. Further, rates of different streams are not always known to the scheduler. Also, these schemes have significant computational complexity, making them relatively difficult to implement in high speed switches.

For completeness, we also mention two other somewhat relevant bodies of work, akin in spirit to our modeling approach. Our ``soft deadline'' point of view discussed before is reminiscent of the time/utility function (TUF) approach introduced by Jensen et. al. \cite {jensen} to study scheduling in real-time operating systems. Moreover, our notion of target profiles for different traffic streams is reminiscent of the rich set of network calculus tools developed by Cruz (\cite {cruz} and several subsequent works with others) to study the problem of providing deterministic QoS guarantees in time-slotted virtual circuit networks, based on the notion of service curves.


\subsection{Contributions}
\label{ssec:contribute}

The key contributions of our work are two-fold. Firstly, we develop a novel {\em outflow aware} switching framework, based on the idea of shaping the switch outflow streams to match desired/target profiles. While we exclusively study this model in the context of an IQ switch, the core ideas are more widely applicable to any queuing system where competing users/jobs are associated with inter-departure time (IDT) constraints.

Secondly, we develop relatively low complexity scheduling policies for IQ switches, using the idea of switch configuration subset based schedules. The idea is to partition the huge set of possible switch service configurations (of size $N!$) into smaller subsets of size $N$ each, and schedule the switch using only one subset in every time-slot. The resulting policies achieve relatively low complexity.
The results presented here provide a substantial extension of the research thread initiated in \cite {globecom,allerton}, where some early observations regarding the studied switch model were made.

In contrast to the previously cited works, in our switch model we do not make any assumptions on the rate, periodicity etc. of traffic streams traversing the switch. We also  develop a family of scheduling policies achieving lower complexity of ${\cal O}(N^2)$ per time-slot, which could be manageable from an implementation point of view in certain practical situations.

\subsection{Organization of the paper}
\label{ssec:outline}

The remainder of this paper is organized as follows: In Section \ref {sec:stc}, we first formulate the service trace control (STC) problem for minimizing stream distortion with respect to their target profiles as a finite-horizon dynamic program \cite {dp}. We then establish the optimality of a {\em greedy policy} for a $2 \times 2$ switch and explore its feasibility as a heuristic policy for bigger switches. Subsequently, we introduce the notion of switch configuration subset based STC in Section \ref {sec:config}. In Section \ref {sec:metaqueue}, we develop the notion of {\it meta-queues}, which yields an alternative view of subset based STC and also provides a general framework for designing different families of STC policies. In Section \ref {sec:admit}, we define the admissible region of the switch and show (using Lyapunov techniques) that subset based STCs, with appropriate subset selection rules, guarantee finite deviation from targets for all traffic streams, under {\it any} admissible load. Experimental evaluation of various proposed scheduling/STC policies in Section \ref {sec:simulation} demonstrates high-performance under various stress regimes. The paper concludes in Section \ref {sec:conclude}.

\subsection{Notations and conventions}
\label{ssec:notation}

Notations and conventions employed throughout the paper are summarized here for convenience. All vectors and sequences are denoted in {\bf boldface}. For a vector $\x$, the $n^{th}$ element is denoted by $x_n$, and for a vector $\x_i$, the $n^{th}$ element is denoted by $x_{i,n}$. $\mathbb{N}$ denotes the set of natural numbers, $\mathbb{Z}$ denotes the set of integers, and $\mathbb{Z}_+$ denotes the set of non-negative integers. $\mathbf{0}$ denotes the all zeros vector and $\mathbf{1}$ denotes the all ones vector. $\mathbf{e}_i$ denotes the $i^{th}$ unit vector in $\mathbb{R}^N$, i.e., a vector with a 1 in the $i^{th}$ location and 0's elsewhere. Further, $\mathbf{e}_0 = \zero$. The inner product between two vectors $\x$ and $\mathbf{y}$ is denoted $\langle \x,\mathbf{y} \rangle$. Finally, the ``big-oh'' notation $f(N) = {\cal O}(g(N))$ is used to indicate that $\exists \; c>0$ such that $f(N) \leq cg(N)$ for large enough $N$.

\section{Minimizing Stream Distortion}
\label{sec:stc}

\subsection{Switching model}
\label{ssec:definition}

Consider an input queued (IQ) switch with virtual output queues (VOQs) at all input ports to prevent head-of-line (HOL) blocking. There are $N^2$ VOQs in an $N \times N$ switch with $N$ input and $N$ output ports, as shown in Fig. \ref {fig:switch}. Both input and output ports are indexed $1,\ldots,N$. The $i^{th}$ VOQ stores packets destined from input port $\lfloor (i-1)/N \rfloor+1$ to output port $(i-1) \mod N + 1$ and is denoted ${\cal Q}_i$. The switch operates in slotted time.
Every input (output) port can be connected to at most one output (input) port in a time-slot. An $N \times N$ switch can be set into $N!$ possible {\it configurations}. Each configuration is associated with a unique {\it configuration vector} of length $N^2$. Let $\v_i = (v_{i,1} \; v_{i,2} \ldots v_{i,N^2}) \in {\cal V}$ denote the $i^{th}$ configuration vector, where ${\cal V}$ is the set of all possible configuration vectors. Then, $v_{i,j}=1$ if ${\cal Q}_j$ is served when the switch is set in configuration $\v_i$ and $v_{i,j}=0$ otherwise. We use the terms configuration and configuration vector interchangeably throughout the paper. 

\begin{numex}
Two possible configuration vectors for a $2 \times 2$ switch are $\v_1 = (1 \; 0 \; 0 \; 1)$ and $\v_2 = (0 \; 1 \; 1 \; 0)$. If a $2 \times 2$ switch is configured with configuration vector $\v_1$, the first (second) input port is connected to the first (second) output port.
If the switch is configured with vector $\v_2$, the first (second) input port is connected to the second (first) output port.
\label{numex:2x2}
\end{numex}

{
\psfrag{VOQs}{VOQs}
\psfrag{Switch fabric}{Switch fabric}
\psfrag{1}{$1$}
\psfrag{2}{$2$}
\psfrag{N}{$N$}
\psfrag{Input Ports}{Input ports}
\psfrag{Output Ports}{Output ports}
\psfrag{STC}{STC}
\begin{figure}
\centerline{\epsfig{file=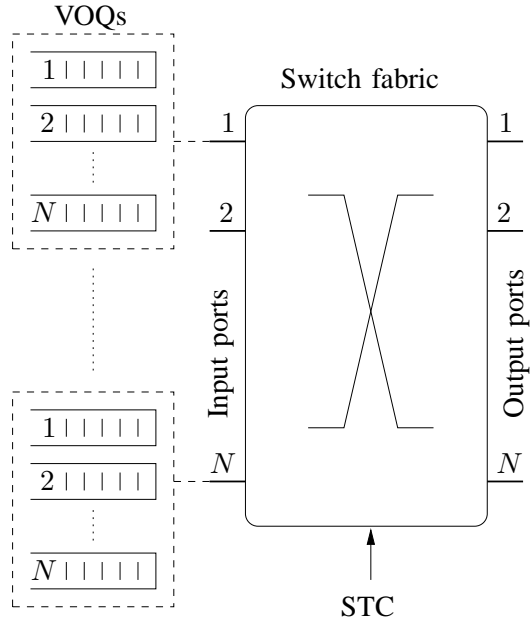,scale=0.75}}
\caption{Input Queued Switch}
\label{fig:switch}
\end{figure}
}

In each time-slot, a single cell can be transferred from an input port to an output port, if those ports are connected in the selected switch configuration. This cell/packet resides in the VOQ associated with the input-output port pair. We use the terms packet and cell interchangeably. Indeed, a cell is a packet of size 1. The underlying assumption is that a packet of size $K$ cells can be ``broken'' into $K$ cells for individual processing, and reassembled at the output of the switch.

We assume there is a large (theoretically infinite) supply of cells/packets residing at the VOQs initially, so that VOQs never run out of packets. For example, one may consider a switch in a video server farm, where video content is retrieved from hard disks and streamed via the switch to remote users. The switch VOQs are directly fed with video packets from the server disk and never (rarely) empty until the streamed content transmission completes. Analogous scenarios emerge in storage area network switches, where large files are retrieved from hard disks and streamed via switches to users.

Every VOQ is associated with a {\it traffic stream}, characterized by a {\bf Target Stream Profile} (TSP). The traffic stream's cells/packets are stored in the associated VOQ. The TSP is the {\it desirable profile of outflow traffic}, i.e., of the stream leaving the switch. It basically specifies the time-slots in which cells of the stream should ideally depart the switch. Alternatively, it characterizes the ideal time distance (number of slots) for releasing two consecutive cells from the stream's VOQ and getting them through and out of the switch.

Technically, the TSP is a sequence of ``0''s and ``1''s which specifies the packet {\it inter-departure time} (IDT) targets/constraints between packets in the stream. Let
\eq
\s = (s^1,s^2,\ldots)
\en
denote the TSP for a typical traffic stream. Suppose that the $k^{th}$ ``1'' in $\s$ occurs at location $\tau \in \mathbb{N}$ and the $(k+1)^{st}$ ``1'' occurs at location $\tau+\delta_k$, for some $\delta_k \in \mathbb{N}$. The interpretation is that the $k^{th}$ packet in the stream should ideally depart the switch in the $\tau^{th}$ time-slot, the $(k+1)^{st}$ packet in the stream should depart the switch in the $(\tau+1)^{st}$ time-slot, and therefore the desired inter-departure time (IDT) target between the $k^{th}$ and $(k+1)^{st}$ packets of the stream is $(\tau+\delta_k)-\tau = \delta_k$ time-slots.
From the TSP we derive the {\bf cumulative Target Stream Profile} (cTSP), denoted $\S = (S^1,S^2,\ldots)$, $S^t \in \mathbb{Z}_+ \; \forall \; t$, where
\eq
\displaystyle S^t \triangleq \sum_{\tau=1}^t s^\tau, \; t = 1, 2, \ldots
\en
is the number of packets of the stream which should ideally have departed the switch by the end of the $t^{th}$ time-slot.

\begin{numex}
We illustrate the concepts of TSP and cTSP through an example. Let the TSP of a stream be given by $\s = (0, 1, 0, 0, 1, 0, 1, \ldots)$. This implies that the $1^{st}$ packet of the stream should ideally depart the switch in the $2^{nd}$ time-slot, the $2^{nd}$ packet should ideally depart in the $5^{th}$ time-slot, the $3^{rd}$ packet should ideally depart in the $7^{th}$ time-slot, and so on. The entries of the cTSP
are computed (by definition) as $S^1 = s^1, S^2 = s^1+s^2, \ldots$. Thus, we have $\S = (0, 1, 1, 1, 2, 2, 3, \ldots)$. The interpretation is that no packets from this stream should have departed the switch by the end of the $1^{st}$ time-slot, exactly one packet should have departed by the end of the $2^{nd}$ time-slot, etc.
\label{numex:basic}
\end{numex}

\begin{numex}
A special example is that of {\bf periodic traffic} (of period $\delta$) with fixed inter-departure times, i.e., $\delta_k = \delta \; \forall \; k$. In this case,  $s^t = 1$ if $t \mod \delta = 0$, and $s^t = 0$ otherwise. Further, $S^t = \lfloor t/\delta \rfloor$. In general, the IDT targets may not be constant but vary substantially, for instance, because of coding dependencies of cell/packets in video streams, etc.
\label{numex:periodic}
\end{numex}

To characterize the service provided by the switch, we associate with every stream a {\bf Received Service Trace} (RST), also a sequence of ``0''s and ``1''s. This is the actual (not desired) service sequence received by the stream.
Let
\eq
\r = (r^1,r^2,\ldots)
\en
denote the RST associated with a typical stream. Then, $r^\tau=1$ if the switch serves a packet from the stream in the $\tau^{th}$ time-slot, and $r^\tau=0$ otherwise. Similar to the cTSP, we derive the {\bf Cumulative Received Service Trace} (cRST), denoted $\R \triangleq (R^1,R^2,\ldots)$, $R^t \in \mathbb{Z}_+ \; \forall \; t$, where
\eq
\displaystyle R^t = \sum_{\tau=1}^t r^\tau, \; t = 1, 2, \ldots
\en
is the number of packets of the stream which have actually departed the switch by the end of $t^{th}$ time-slot.
\begin{numex}
We illustrate the concepts of RST and cRST through an example. Consider a traffic stream with its TSP as given by Example \ref {numex:basic}. Now, suppose the RST for this stream is given by $\r = (0, 1, 0, 1, 0, 0, 0, 1, \ldots)$. The interpretation is that the $1^{st}$ packet of the stream departed the switch in the $2^{nd}$ time-slot, the $2^{nd}$ packet departed in the $4^{th}$ time-slot, the $3^{rd}$ packet departed in the $8^{th}$ time-slot, and so on. By definition, the cRST is constructed as $R^1 = r^1, R^2 = r^1+r^2, \ldots$, yielding $\R = (0, 1, 1, 2, 2, 2, 2, 3, \ldots)$. The interpretation is that no packets from the stream were released from the switch by the end of the $1^{st}$ time-slot, one packet was released by the end of the $2^{nd}$ time-slot, etc.
\label{numex:basic2}
\end{numex}

{
\psfrag{Time}{Time}
\psfrag{rs}{cRST}
\psfrag{es}{cTSP}
\psfrag{lead}{Lead}
\psfrag{lag}{Lag}
\psfrag{Cost}{Cost}
\psfrag{lagcost}{Lag Cost}
\psfrag{leadcost}{Lead Cost}
\psfrag{deviation}{Deviation}
\begin{figure*}[t]
\centerline{\epsfig{file=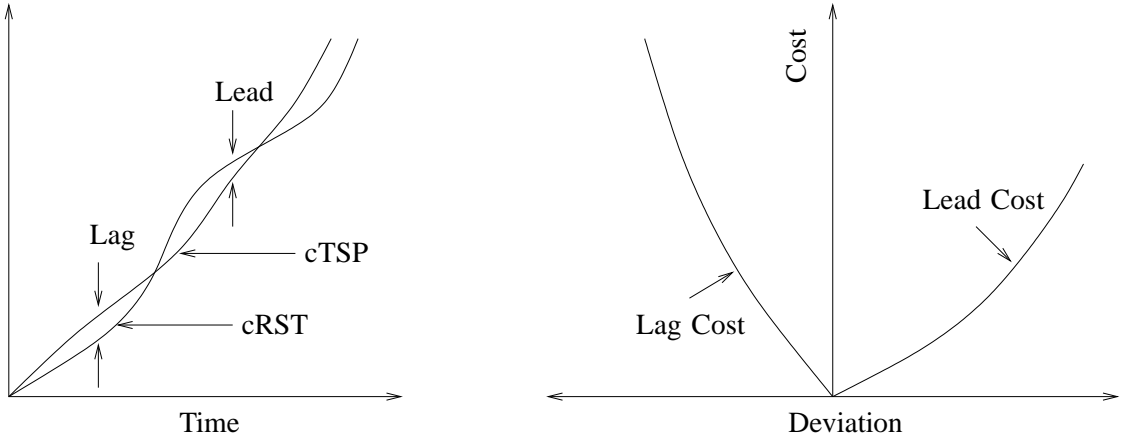,scale=0.75}}
\caption{The left side illustrates the notions of cTSP, cRST and deviation. The right side depicts a typical deviation cost function.}
\label{fig:deviation}
\end{figure*}
}

Ideally, for every stream we desire $R^t = S^t \; \forall \; t$, which implies that every stream traverses the switch without experiencing any ``distortion'' of its target profile. However, this goal is not always realizable due to congestion caused by contention between competing streams for the shared switch fabric. If for a particular stream $R^t > S^t$, the stream has received more service than it requires to satisfy its packet inter-departure time (IDT) constraints and is said to be {\bf leading} at time $t$. If $R^t < S^t$, the stream has received less than its desired amount of service and is said to be {\bf lagging} at time $t$. To quantify distortion of target profiles due to congestion, we track for every traffic stream its {\bf deviation}, denoted $\d \triangleq (d^1, d^2, \ldots)$, $d^t \in \mathbb{Z} \; \forall \; t$, where
\eq
\label{devdev}
d^t \triangleq R^t-S^t, \; t = 1, 2, \ldots
\en
which quantifies the excess or deficiency in service catered to the stream by the switch as a function of time. A negative deviation ({\bf lag}) indicates missed deadlines and is undesirable from a QoS provisioning perspective. A positive deviation ({\bf lead}) is undesirable because it can cause buffer overflows at downstream switches and the end user and lead to starvation of delay tolerant flows traversing the switch.

\begin{numex}
We illustrate the notion of deviation through an example. Consider a traffic stream with its TSP as given by Example \ref {numex:basic} and RST as given by Example \ref {numex:basic2}. Recall that for this stream, the cTSP is given by $\S = (0, 1, 1, 1, 2, 2, 3, \ldots)$ and the cRST is given by $\R = (0, 1, 1, 2, 2, 2, 2, 3, \ldots)$. Taking an elementwise difference, the deviation is given by $\d = \R - \S = (0, 0, 0, 1, 0, 0, -1, \ldots)$. Note that the $1^{st}$ packet of the stream gets served by the switch on time, the $2^{nd}$ packet gets served one time-slot in advance, and the $3^{rd}$ packet gets served one time-slot later than desired. The stream is therefore ``leading'' for one time-slot immediately after the departure of the $2^{nd}$ packet, and is lagging in the $7^{th}$ time-slot, which is the desired/target departure time of the $3^{rd}$ packet in the stream.
\label{numex:basic3}
\end{numex}
The ideas introduced in this section are depicted in Fig. \ref {fig:deviation}. While the cTSP and cRST curves are shown to be ``smooth'' in the figure for illustration, note that for the discrete-time model studied in this paper (at most one cell per VOQ processed by the switch in a time-slot), the curves will look like staircase functions, with the step size equal to 1.

\subsection{Finite horizon dynamic programming (DP) formulation}
\label{ssec:dpform}

Consider a finite horizon of $T>0$ time-slots indexed by $t \in \{1,\ldots,T\}$. Let $\s_i = (s_i^1,\ldots,s_i^T)$  and $\S_i = (S_i^1,\ldots,S_i^T)$  denote the the first $T$ entries of the TSP and cTSP of VOQ ${\cal Q}_i$, respectively.  Define
\eq
\label{aaabbb}
\x^t \triangleq (s_1^t,\ldots,s_{N^2}^t), \; X^t \triangleq (S_1^t,\ldots,S_{N^2}^t).
\en
Thus, $\x^t$ ($\X^t$) is a vector of the $t^{th}$ entries of the TSP (cTSP) of all the $N^2$ streams. We shift from the $\s_i$ ($S_i$) notation to the $x^t$ ($X^t$) one in order to change the point of view from being focused on each individual queue/stream $i$ to tracking (all queues/streams at) each time-slot $t\leq T$. To clarify the notation further, consider a $T \times N^2$ matrix with $\x^1, \x^2, \ldots, \x^t$ as its rows. The $i^{th}$ column of this matrix comprises of the TSP entries for ${\cal Q}_i$ over the time horizon of interest, viz. $\{1,\ldots,T\}$, In matrix terminology, the $i^{th}$ column is the transpose of the TSP of ${\cal Q}_i$. On the other hand, the $t^{th}$ row comprises of the $t^{th}$ entries of all $N^2$ traffic streams traversing the switch.
Next, define
\eq
\d^t \triangleq (d_1^t,\ldots,d_{N^2}^t)
\en
as the {\bf state} of the switch in the $t^{th}$ time-slot, where $d_i^t$ is the {\em deviation} (as defined in (\ref{devdev})) of the stream associated with VOQ ${\cal Q}_i$ in the $t^{th}$ time-slot.

Since deviations from target profiles are undesirable, they are associated with a ``cost''. In particular, to the $i^{th}$ stream we assign the cost function $\phi_i(k)$, which reflects the cost associated with a deviation $k \in \mathbb{Z}$. We assume the following:
\begin{enumerate}
\item $\phi_i(0)=0$ (zero deviation is desirable)
\item $\phi_i(k)$ is non-negative and increasing for both $k>0$ and $k<0$ (since both leads and lags are undesirable)
\item $\phi_i(k)$ is convex (the cost associated with deviation increases at a positive rate as the deviation increases in magnitude) 
\end{enumerate}
A sample cost function which satisfies the above properties is depicted on the right side of Fig. \ref {fig:deviation}. 
An example of a cost function which we will often use in this paper is the quadratic cost function $\phi_i(k) = k^2$. Finally, let
\eq
\displaystyle \Phi(\d^t) \triangleq \sum_{i=1}^{N^2} \phi_i(d_i^t)
\label{Phidef}
\en
denote the sum of the deviation costs of all VOQs.

\begin{remarks}
It is important to note that unlike the packet inter-departure time constraints, the cost functions are not an inherent part of the problem, but are instead extraneously assigned by the switch controller for the purpose of service trace control. Thus, the switch controller has the freedom to {\em tune} these cost functions in order to optimize switch performance.
\end{remarks}

\begin{remarks}
In our modeling framework, packets can be thought of as being associated with {\em soft deadlines}. For the more conventional case of strict deadlines, the ``value'' of a packet is constant prior to its deadline and zero thereafter. As a result, a packet is dropped if it has not departed the queue before its due date. In our context, where a typical motivating application is multimedia streaming, lossless delivery of packets is sought. The ``value'' of a packet reaches its peak at its target delivery time (as dictated by the TSP). The packet is treated as less valuable (but not dropped) if received either before or after its target time. In this sense, the ``softness'' of the deadline constraints for a traffic stream is quantified by the steepness of the associated cost function.
\end{remarks}

In every time-slot, the {\bf Service Trace Controller} (STC) drives the evolution of service traces for various traffic streams by setting the switch in one of $N!$ possible configurations (chosen from the set ${\cal V}$)
or idling the switch.

\begin{definition}
A {\bf policy} $\Pi_T=\{\v^t \in {\cal V} \cup \{\zero\}, \; t=1,\ldots,T\}$ is defined as a sequence of switch configurations selected by the service trace controller in time-slots $t=1,\ldots,T$. 
\end{definition}

Given the initial state $\d^0$, we are interested in computing the optimal policy (one which minimizes the total cost over a finite horizon) $\Pi_T^\star$ which satisfies
\begin{equation}
\Pi_T^\star = \mathop{\arg \min}_{\Pi_T}  \left\{ \sum_{t=1}^T \Phi(\d^t_{\Pi_T}) \right\},
\label{pistar}
\end{equation}
where $\d^t_{\Pi_T}$ denotes the state of the switch at the beginning of the $t^{th}$ time-slot under policy $\Pi_T$. We will adopt the methodology of {\it dynamic programming} (DP) to compute $\Pi_T^\star$.

Suppose $\Pi_T$ chooses configuration vector $\v^t=\v = (v_1, v_2, \ldots, v_{N^2})$ in the $t^{th}$ time-slot. The deviation of the VOQ ${\cal Q}_i$ increases by 1 at the end of the $t^{th}$ time-slot if it is served by configuration $\v$, i.e., $v_i=1$. Also, the deviation of the ${\cal Q}_i$ decreases by 1 if its TSP has a non-zero entry in the $t^{th}$ location, i.e., $x_i^t=1$. Note that $x_i^t$ (the $i^{th}$ component of $\x^t$) is simply $s_i^t$, from (\ref{aaabbb}).
More compactly, the new deviation vector at the beginning of the $(t+1)^{st}$ time-slot is given by
\eq
\d^{t+1}_{\Pi_T} = \d^t_{\Pi_T} + \v^t - \x^t.
\en

Let $V^t(\d)$ be the cost incurred by $\Pi_T^\star$ over time-slots $t,\ldots,T$, starting in state $\d$ at the beginning of the $t^{th}$ time-slot. In dynamic programming terminology, $V^t(\cdot)$ is referred to the as the {\it cost-to-go} function, and is recursively computed from the following DP equations for $t=1,\ldots,T$
\begin{equation}
V^t(\d) = \mathop{\min}_{\v \in {\cal V} \cup \{\zero\}} \left\{ \underbrace{V^{t+1}(\d + \v - \x^t)}_{\text{Cost-to-go in the next time-slot}} + \underbrace{\Phi(\d + \v - \x^t)}_{\text{Instantaneous cost}} \right\},
\label{dp1}
\end{equation}
and the boundary conditions $V^{T+1}(\d)=0 \; \forall \; \d$. We will henceforth refer to $\d-\x^t$ as the {\bf deviation vector}.

\subsection{Myopic/Greedy service trace control}
\label{ssec:myopic}

Observe from (\ref {dp1}) that the optimal decision in state $\d$ in the $t^{th}$ time-slot is determined by the cost-to-go in the $(t+1)^{st}$ time-slot, as well as the instantaneous cost. Now consider a {\it myopic policy}, which is ``greedy'' with respect to the instantaneous cost, i.e., ignores the cost-to-go in the next time-slot while making its current scheduling decision. In particular, the myopic policy chooses configuration $\v^\star$ in the $t^{th}$ time-slot in state $\d$ such that
\begin{equation}
\v^\star = \mathop{\arg \min}_{\v \in {\cal V} \cup \{\zero\}} \left\{ \Phi(\d + \v - \x^t) \right\}.
\label{myopic}
\end{equation}
In general, the myopic policy need not be optimal. However, for the scheduling problem at hand, the myopic policy is provably optimal for the case $N=2$. For $N=3, 4$, numerical analysis reveals that the myopic policy is close to optimal. The cost of computing the optimal policy becomes prohibitive as $N$ gets bigger ($N!+1$ possible decisions need to be evaluated in every possible state of the switch over a period of $T$ time-slots). 

\vspace{0.05in}

\begin{theorem}
The optimal finite horizon policy $\Pi_T^\star$ for a $2 \times 2$ switch ($N=2$) is myopic.
\label{theorem:2x2myopic}
\end{theorem}

\vspace{0.05in}

\noindent\begin{proof}
See Appendix \ref {ssec:proof1}.
\end{proof}

\vspace{0.05in}

\begin{numex}
\label{numex:qcf}
For concreteness and as a key example, let us assign quadratic cost functions to all traffic streams, i.e., $\phi_i(k)=k^2 \; \forall \; i$. For any $\v \in {\cal V} \cup \{\zero\}$ we have
\begin{equation}
\Phi(\d+\v-\x^t) = \underbrace{\langle \d-\x^t,\d-\x^t \rangle}_{\text{Policy independent}} + \underbrace{\langle \v,\v \rangle + 2 \langle \d-\x^t, \v \rangle}_{\text{Policy dependent : } \varsigma(\v)}.
\label{policydep}
\end{equation}
The myopic policy in this case reduces to
\begin{equation}
\v^\star = \left\{
\begin{array}{ccc}
\widetilde{\v}^\star & ; & 2 \langle \d-\x^t, \widetilde{\v}^\star \rangle + N \leq 0 \\
\zero & ; & \text{else},
\end{array}
\right.
\label{maxproj}
\end{equation}
where $ \displaystyle \widetilde{\v}^\star \triangleq \mathop{\arg \min}_{\v \in {\cal V}} \left\{ \langle \d-\x^t, \v \rangle \right\}$.

The idea is as follows: We want to find a $\v \in \cal V$ which minimizes the policy dependent part. The set $\cal V$ is the set of all $N^2$ switch configuration plus the zero configuration (switch idle). For $\v = 0$, the policy dependent part is 0. For all non-zero configurations, $\langle \v,\v \rangle = N$, and the policy dependent part is $2 \langle \d-\x^t, \v \rangle + N$. This term is minimized by $\widetilde{\v}^\star$ (by definition). Thus, we pick the $\v$ which minimizes the min of 0 and $2 \langle \d-\x^t, \widetilde{\v}^\star \rangle$.
\label{numex:quadratic}
\end{numex}

\subsection{Partial configurations}
\label{ssec:partial}

We begin this section with a definition. 
\begin{definition}
For an $N \times N$ IQ switch, a switch configuration $\v \in {\cal V}$ is called {\bf complete} if $\langle \v, \one \rangle = N$ and is called {\bf partial} if $\langle \v, \one \rangle \leq N$.
\end{definition}

In other words, in a complete configuration, {\it every} input port is connected to an output port, while in a partial configuration, some of the input ports may be idle.

So far we have assumed that the service trace controller (STC) either selects a {\it complete} configuration ({\it every} input port is connected to an output port) 
or idles the switch. However, operating the switch using complete configurations only is not sufficient to exercise individual control on service traces of different streams, as illustrated by the next example.

\begin{numex}
Consider a $2 \times 2$ switch where the streams for ${\cal Q}_1$ and ${\cal Q}_4$ are periodic with periods 2 and 4 respectively, and ${\cal Q}_2$ and ${\cal Q}_3$ are empty (no new arrivals). In our notation, this translates to $\x^1 = (0,0,0,0), \x^2 = (1,0,0,0), \x^3 = (0,0,0,0), \x^4 = (1,0,0,1)$ and $\x^t = \x^{t \mod 4} \; \forall \; t$. The myopic policy (which is optimal) given by (\ref {maxproj}) either selects $\v_1 = (1 \; 0 \; 0 \; 1)$ or $\zero$ in every time-slot. The configuration vector $\v_2$ is never selected because both queues serviced by $\v_2$ are empty. It is easily verified that either the lag of ${\cal Q}_1$ or the lead of ${\cal Q}_4$ grow without bound under the optimal policy. 
\end{numex}

To exercise individual control over service traces, we allow the STC to use {\it partial} configurations. Suppose complete configuration $\v \in {\cal V}$ serves VOQs indexed by the set ${\cal I} = \{i_1,\ldots,i_N\}$. Any partial configuration $\bar{\v}$ extracted from $\v$ is characterized by a vector $\xib = (\xi_1,\ldots,\xi_N)$, where $\xi_j=1$ if $\bar{\v}$ serves ${\cal Q}_{i_j}$ and $\xi_j=0$ if $\bar{\v}$ idles ${\cal Q}_{i_j}$. Thus, $2^N$ partial configurations can be extracted from any complete configuration. 
\begin{numex}
Consider configuration $\v_1 =(1 \; 0 \; 0 \; 1)$ for a $2 \times 2$ switch. This configuration serves VOQs ${\cal Q}_1$ and ${\cal Q}_4$. The partial configuration set $\{(1 \; 0 \; 0 \; 0), (0 \; 0 \; 0 \; 1), \zero,\v_1 \}$ can be extracted from the complete configuration $\v_1$. The first partial configuration in the set corresponds to $\xib = (1,0)$, the second partial configuration corresponds to $\xib = (0,1)$, etc. Note that the configurations $\zero$ and $\v$ are always part of the configuration set associated with complete configuration $\v$.
\end{numex}

\subsection{The Maximum Sum of Lags (MSL) policy}
\label{ssec:msl}
Let us revisit the myopic service trace control policy for the case of quadratic cost functions (Example \ref {numex:quadratic}), allowing for partial configurations this time. Recall from (\ref {policydep}) that the policy dependent part in $\Phi$ is $\varsigma(\v) = \langle \v,\v \rangle + 2 \langle \d-\x^t,\v \rangle$. If $\v$ serves VOQs indexed by set ${\cal I}$, $\varsigma(\v)$ can be rewritten as $\displaystyle \sum_{j \in {\cal I}} v_{j}^2  + 2\sum_{j \in {\cal I}} v_{j}(d_j-x_j^t)$. Since $\v$ is a complete configuration, $v_{j}=1 \; \forall \; j \in {\cal I}$. Now, split ${\cal I}$ into two disjoint subsets, ${\cal I}_+$ and ${\cal I}_-$, where ${\cal I}_+ = \{j \in {\cal I}: d_j-x_j^t \geq 0\}$ and ${\cal I}_- = \{j \in {\cal I}: d_j-x_j^t < 0\}$. Note that ${\cal I}_+ \cup {\cal I}_- = {\cal I}$ and ${\cal I}_+ \cap {\cal I}_- = \emptyset$.  Clearly, $\varsigma(\v)$ can be strictly decreased by setting $v_{j}=0 \; \forall \; j \in {\cal I}_+$. Doing so is equivalent to extracting a partial configuration from $\v$ by idling all VOQs with non-negative deviation. We therefore get the following two-step service trace control policy, which we refer to as the {\bf Maximum Sum of Lags} (MSL) policy (see Table \ref {table:policy}).
\begin{enumerate}
\item Select $ \displaystyle \widetilde{\v}^\star = \mathop{\arg \min}_{\v \in {\cal V}} \left\{ \langle \d-\x^t, \v \rangle \right\}$.
\item Extract a partial configuration from $\widetilde{\v}^\star$ by idling all VOQs with non-negative deviation.
\end{enumerate}
The name of the policy arises from the fact that it selects the switch configuration whose associated VOQs have the largest sum ``lag'' (as defined in Section \ref {ssec:definition}).

The computational complexity of MSL is ${\cal O}(N^3)$ per time-slot, since Step 1 involves a maximum weight matching (MWM) computation on a bipartite graph \cite {clrs}. Note that the edge weights used to compute this matching are in fact the deviations associated with the VOQs. Switching algorithms which use VOQ backlogs as the edge weights for computing MWM have been studied extensively in the literature, in the context of throughput maximizing switches (e.g. \cite {nick1}). 
While ${\cal O}(N^3)$ complexity is a significant improvement over the optimal policy, algorithms to compute the maximum weight matching are cumbersome to implement and impractical for large switches. This motivates us to explore service trace control policies which yield MSL-like performance at manageable complexity.

\begin{remarks}
Step 2 of MSL can be generalized to construct a broader class of policies, namely MSL($\ell$), indexed by $\ell \in \mathbb{N} \cup \{0\}$. Under MSL($\ell$)\footnote{Note that it may not be feasible to realize MSL($\ell$) for arbitrary $\ell > 0$, if the switch cannot provide a lead of $\ell$ even in the absence of congestion, due to unavailability of packets ahead of their departure times. However, MSL($\ell$) is pertinent in a scenario where the switch resides at the egress of a multimedia server, where all traffic streams are pre-cached at the input of the switch. In this case, the switch can furnish a lead of up to $\ell$ to provide a ``cushion'' against possible congestion in the downstream network.}, a VOQ served by the chosen complete configuration is idled only when its deviation is $\ell$ or more. By this token, MSL $\equiv$ MSL(0). 
\label{remarks:mslL}
\end{remarks}

\section{Subset Based Service Trace Control}
\label{sec:config}

To address the issue of high computational complexity associated with optimal service trace control, we propose a {\em subset based control} approach in this section. The key idea is to partition the configuration set ${\cal V}$ of size $N!$ into smaller disjoint subsets of size $N$ each and operate the switch using configurations from only one of these subsets in any time-slot.

\subsection{Subset construction}
\label{ssec:subsets}

It follows by design that all configuration vectors for an IQ switch are of the form $\v = [\e_{\pi(1)} \; \e_{\pi(2)} \ldots \e_{\pi(N)}]$, where $\pi$ is a permutation of $\{1,2,\ldots,N\}$ and $\e_i$ is as defined in Section \ref {ssec:notation}. Now, we define the {\it circular shift operator}. 

\begin{definition}
The {\bf circular shift operator} ${\cal C} : {\cal V} \mapsto {\cal V}$ is given by
\begin{equation}
{\cal C}(\v) \triangleq [\e_{\pi(N)} \; \e_{\pi(1)} \; \e_{\pi(2)} \ldots \e_{\pi(N-1)}],
\end{equation} 
where $\v = [\e_{\pi(1)} \; \e_{\pi(2)} \ldots \e_{\pi(N)}] \in {\cal V}$ is a switch configuration vector.
\end{definition}

Recursively define ${\cal C}^k(\v) \triangleq {\cal C}({\cal C}^{k-1}(\v))$, $k \in \mathbb{N}$, which corresponds to applying the circular shift operator $k$ times to $\v$. By convention, ${\cal C}^0(\v) = \v$. Also, note that ${\cal C}^k(\v) = {\cal C}^{k \mod N}(\v)$. Thus, starting with any configuration vector $\v \in {\cal V}$, we can generate a set of $N$ distinct configuration vectors by applying the operator ${\cal C}$ to $\v$ $N-1$ times. We say that $\v$ {\it generates} the {\it configuration subset}
\eq
{\cal S}_{\v} = \{\v,{\cal C}(\v),\ldots,{\cal C}^{N-1}(\v)\} \subset {\cal V}
\en
and refer to $\v$ as the {\it generator vector}. Following the outlined procedure, we can partition ${\cal V}$ into $(N-1)!$ disjoint configuration subsets of size $N$ each. As an example, the configuration subsets for a $3 \times 3$ switch are depicted in Fig. \ref {fig:ortho}.

For any $\v \in {\cal V}$, $\langle \v, {\cal C}^k(\v) \rangle=0$ $\forall \; k \in \mathbb{N}$. Physically, this implies that no VOQ is served by more than one configuration in a subset. Geometrically, this mean that all configuration vectors within a subset are ``orthogonal'' to each other. We therefore say that the generated subsets are {\it orthogonal}. Also, note that every VOQ is served by some configuration within a subset. Consequently, we say that every subset is {\it complete}. Combining the orthogonality and completeness properties we see that every VOQ is associated with {\it exactly} one configuration vector in every subset, implying
\begin{equation}
\sum_{j=0}^{N-1} {\cal C}^j(\v) = \one \quad \forall \; \v \in {\cal V}.
\label{subset}
\end{equation}

{
\psfrag{x}{$\v_1$}
\psfrag{cx}{${\cal C}(\v_1)$}
\psfrag{c2x}{${\cal C}^2(\v_1)$}
\psfrag{y}{$\v_2$}
\psfrag{cy}{${\cal C}(\v_2)$}
\psfrag{c2y}{${\cal C}^2(\v_2)$}
\begin{figure*}[t]
\centerline{\epsfig{file=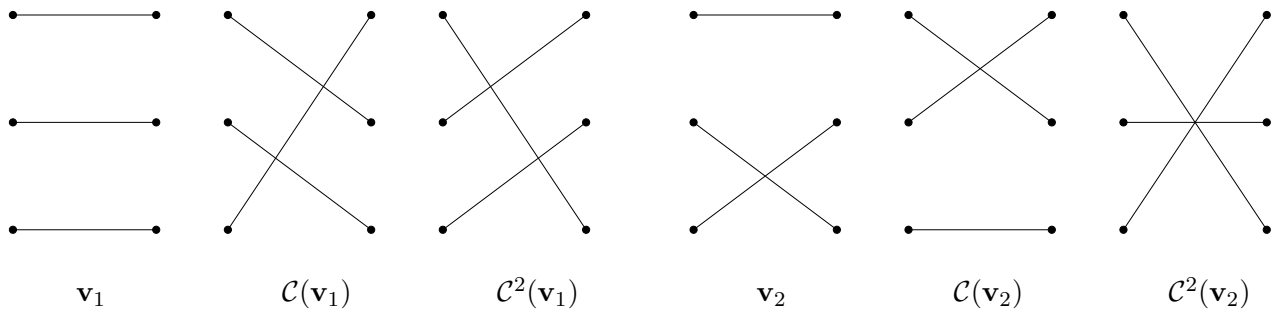,scale=0.75}}
\caption{Two configuration subsets (of size 3 each) for a $3 \times 3$ switch. The three leftmost configurations are generated by $\v_1=[\e_1 \; \e_2 \; \e_3]$ and the three rightmost configurations are generated by $\v_2=[\e_1 \; \e_3 \; \e_2]$.}
\label{fig:ortho}
\end{figure*}
}

\subsection{The MSL-SS policy}
\label{ssec:msls}

Let ${\cal S}_\v=\{{\cal C}^i(\v)\}_{i=0}^{N-1}$ be the configuration subset generated by $\v$. Now consider operating the switch such that the service trace controller is allowed to choose configurations from ${\cal S}_\v$ alone, rather than from ${\cal V}$. In particular, consider a restriction of the MSL policy of Section \ref {ssec:msl} to the configuration subset ${\cal S}_\v$. We get the following two-step policy, which we call the {\bf Maximum Sum of Lags - Single Subset} (MSL-SS) policy:
\begin{enumerate}
\item Select configuration ${\cal C}^{i^\star}(\v) \in {\cal S}_\v$ such that
\begin{equation}
i^\star = \mathop{\arg \min}_{i=0,\ldots,N-1} \left\{ \langle \d-\x^t, {\cal C}^i(\v) \rangle \right\}.
\end{equation}
\item Extract a partial configuration from ${\cal C}^{i^\star}(\v)$ by idling all VOQs with non-negative deviation. 
\end{enumerate}
The per time-slot computational complexity for MSL-SS is ${\cal O}(N^2)$, in contrast to ${\cal O}(N^3)$ for MSL. 

\begin{remarks}
To compute the optimal decision for MSL-SS, $N$ inner products of the form $\langle \d-\x^t, {\cal C}^i(\v) \rangle$ need to be computed, followed by a $\min$ of the resulting $N$ numbers. Each of these inner products involve vectors of length $N^2$. However, note that one of the vectors involved in each inner product is a configuration vector, which is relatively sparse (only $N$ out of the $N^2$ entries are non-zero). Further, all the non-zero entries are equal to 1. Each inner product, $\langle \d-\x^t, {\cal C}^i(\v) \rangle$, is therefore simply a sum of $N$ numbers.
The MSL-SS policy is thus straightforward to implement, compared to algorithms used for computing maximum weight matching (needed for MSL). 
\end{remarks}
Two crucial questions arise at this point:
\begin{enumerate}
\item What is the performance loss (if any) incurred by operating the switch using only one configuration subset?
\item Can we compensate for the loss (if needed), without sacrificing the advantage of low complexity?
\end{enumerate}
We will address these questions in the remainder of the paper.

\section{Meta-Queue Based Service Trace Control}
\label{sec:metaqueue}

In this section, we study subset based service trace control in a broader framework, based on the notion of meta-queues. We will recover the MSL-SS policy proposed in Section \ref {ssec:msls} as a special case of the meta-queue framework.

\subsection{Meta-queue construction}
\label{ssec:mq}

Setting the switch in the complete configuration given by $\v=[\e_{\pi(1)}  \ldots \e_{\pi(N)}]$ is equivalent to serving VOQs indexed by the set ${\cal I}=\{(i-1)N + \pi(i), \;  i=1,\ldots,N\}$. Thus, every complete configuration serves $N$ VOQs concurrently, which we ``group'' together to form a {\it meta-queue}.

Let us focus on a single subset, say ${\cal S}_\v$. Since ${\cal S}_\v$ is orthogonal and complete by construction,
each configuration in ${\cal S}_\v$ can be associated with a unique meta-queue, constructed by ``grouping" $N$ distinct VOQs. Note that all $N^2$ VOQs are assigned to some meta-queue, each one exactly once. The head of line (HOL) {\it meta-packet} of a meta-queue is constructed by grouping the HOL packets of its $N$ constituent VOQs. 
With this construction, choosing a switch configuration is equivalent to serving the HOL meta-packet of the corresponding meta-queue.

While grouping concurrently served VOQs to form a meta-queue seems quite natural, the relation between the deviation of a meta-queue and the deviations of its constituent VOQs is not immediately evident. In fact, we have the freedom to choose a mapping $\Gamma: \mathbb{Z}^N \mapsto \mathbb{Z}$, which relates the deviation of a meta-queue to the deviations of its $N$ constituent VOQs. Given a mapping $\Gamma$, the problem of subset based control of an IQ switch turns into a problem of scheduling $N$ parallel meta-queues on a single server. The latter is an important and interesting scheduling problem in its own right (e.g. see \cite {walrand}).

We now briefly digress from the service trace control problem for the IQ switch to study the single server scheduling problem mentioned above. Subsequently, we will show that by appropriately choosing $\Gamma$, one can construct good, low complexity service trace control policies for an IQ switch.

\subsection{The single server scheduling problem}
\label{ssec:ss}

The formulation is similar in spirit to the formulation for an IQ switch (Section \ref {ssec:dpform}), and so is the notation. Consider a system comprised of $N+1$ parallel meta-queues and a single server. The $i^{th}$ meta-queue is denoted ${\cal M}_i$, $i=0,\ldots,N$. In every time-slot the scheduler serves the HOL meta-packet of one of the meta-queues, chosen according to some scheduling policy. While ${\cal M}_1,\ldots,{\cal M}_N$ are ``physical'' meta-queues, ${\cal M}_0$ is a ``dummy'' meta-queue, scheduling which is tantamount to idling the server. 
Each meta-queue is associated with a traffic stream characterized by a target service profile (TSP). The interpretation of the TSP in this context is identical to Section \ref {ssec:definition}, i.e., it specifies the time-slots in which meta-packets from a meta-queue should ideally depart the server. The TSP associated with ${\cal M}_0$ has all zero entries. We denote by $\dwb^t = (\dw_1^t,\ldots, \dw_N^t)$ the deviation vector for the system in the $t^{th}$ time-slot, where $\dw_i^t$ is the deviation for ${\cal M}_i$. Define $\xwb^t \triangleq (\sw_1^t,\ldots,\sw_N^t)$ and $\Xwb^t \triangleq (\Sw_1^t,\ldots,\Sw_N^t)$, where $\sw_i^t$ and $\Sw_i^t$ are respectively the $t^{th}$ elements of the TSP and cumulative TSP (cTSP) of ${\cal M}_i$. To ${\cal M}_i$ we assign the cost function $\psi_i(k)$, which quantifies the cost of deviation $k \in \mathbb{Z}$. Similar to Section \ref {ssec:dpform}, we assume that $\psi_i(k)$ is non-negative, convex, and increasing for both $k>0$ and $k<0$, and $\psi_0(k) = 0 \; \forall \; k$. Finally, let
\eq
\displaystyle \Psi(\d^t) \triangleq \sum_{i=1}^N \psi_i(d_i^t)
\en
denote the sum of deviation costs of all meta-queues.

We confine our attention to a finite horizon of $T$ time-slots. At the beginning of every time-slot, the scheduler selects one of the $N+1$ meta-queues for service. The configuration vector corresponding to scheduling ${\cal M}_i$ is $\e_i$. An admissible policy $\widetilde{\Pi}_T$ for the single server scheduling problem is a sequence of scheduling decisions $\{i_t\}_{t=1}^T$, corresponding to scheduling meta-queue ${\cal M}_{i_t}$ in the $t^{th}$ time-slot. Let $\dwb^t_{\widetilde{\Pi}_T}$ denote the deviation vector at the beginning of the $t^{th}$ time-slot under scheduling policy $\widetilde{\Pi}_T$. 
Our goal is to compute the optimal finite horizon policy $\widetilde{\Pi}_T^\star$ which satisfies
\begin{equation}
\widetilde{\Pi}_T^\star = \mathop{\arg \min}_{\widetilde{\Pi}_T} \left\{ \sum_{t=1}^T \Psi(\dwb^t_{\widetilde{\Pi}_T}) \right\}.
\label{opt}
\end{equation}
We specify the state of the system at the end of the $t^{th}$ time-slot by
\eq
\n^t = (n_1^t,\ldots,n_N^t),
\en
where $n_i^t$ is the number of times ${\cal M}_i$ has been served within the first $t$ time-slots. Since the server is allowed to idle, $\langle \n^t, \one \rangle \leq t \; \forall \; t$. The system state and deviation vector are uniquely related by
\eq
\dwb^{t+1} = \dwb^0 + \n^t - \Xwb^t.
\en
If the state at the beginning of the $t^{th}$ time-slot is $\n$ and the scheduler chooses ${\cal M}_i$ in the $t^{th}$ time-slot, the new state at the beginning of the $(t+1)^{st}$ time-slot is $\n+\e_i$. Letting $\widetilde{V}^t(\n)$ denote the cost-to-go at the beginning of the $t^{th}$ time-slot in state $\n$, we have the following DP equations for $t=1,\ldots,T$
\begin{equation}
\widetilde{V}^t(\n) = \mathop{\min}_{i=0,\ldots,N} \left\{ \widetilde{V}^{t+1}(\underbrace{\n+\e_i}_{\text{New state}}) + \Psi(\underbrace{\n+\e_i-\Xwb^t}_{\text{New deviation vector}}) \right\},
\label{dpeqnss}
\end{equation}
and the boundary conditions $\widetilde{V}^{T+1}(\n) = 0 \; \forall \; \n$. For notational convenience, define
\eq
\Omega^t(\n) \triangleq \widetilde{V}^t(\n) + \Psi(\n-\X^t).
\en
Also, define the {\bf pairwise decision functions}
\begin{equation}
\gamma_{ij}^t(\n) \triangleq \Omega^{t+1}(\n+\e_i) - \Omega^{t+1}(\n+\e_j), \quad i \neq j.
\label{gammadef}
\end{equation}
It follows that $\widetilde{\Pi}_T^\star$ ``prefers'' ${\cal M}_i$ over ${\cal M}_j$ in the $t^{th}$ time-slot in state $\n$ if $\gamma_{ij}^t(\n) \leq 0$, and ``prefers'' ${\cal M}_j$ else. The pairwise decision functions satisfy the following:

\vspace{0.05in}

\begin{lemma}[Monotonicity of $\gamma$]
$\gamma_{ij}^t(\n)$ is a non-decreasing function of $n_i$ and a non-increasing function of $n_j$ for $i,j \in \{0,\ldots,N\}$, $i \neq j$, and $t=1,\ldots,T$.
\label{lemma:gammalemma}
\end{lemma}

\vspace{0.05in}

\begin{proof}
See Appendix \ref {ssec:gammalemmaproof}.
\end{proof}

\vspace{0.05in}

Lemma \ref {lemma:gammalemma} can be used to show that any two-dimensional subspace of the $N$-dimensional state-space is partitioned into $N+1$ connected {\it decision regions} by $\widetilde{\Pi}_T^\star$. The states in the $i^{th}$ decision region are those in which $\widetilde{\Pi}_T^\star$ schedules ${\cal M}_i$ in the $t^{th}$ time-slot. Further, for every $t$, as $n_i^t$ increases for fixed $n_j^t$, $j \neq i$, $\widetilde{\Pi}_T^\star$ {\it switches over} from ${\cal M}_i$ to ${\cal M}_k$ for some $k \neq i$. Thereafter, $\widetilde{\Pi}_T^\star$ {\it never} switches back to ${\cal M}_i$. Unfortunately, this neat structural insight does not immediately yield a low complexity approximation of the optimal policy.

\begin{figure}[ht]
\centerline{
\epsfig{file=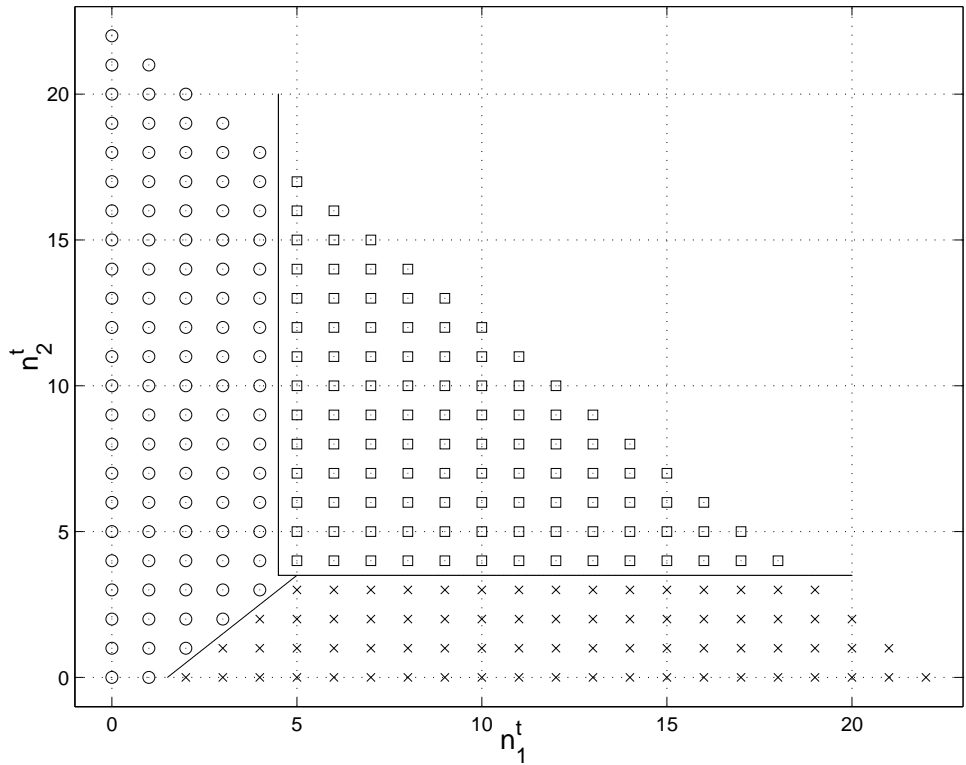,scale=0.75}}
\caption{Partitioning of the $(n_1^t,n_2^t)$ plane into decision regions for fixed $n_3^t=8,T=40,t=30$ for $p=0.1$. The states in which it is optimal to schedule ${\cal M}_0, {\cal M}_1, {\cal M}_2$, and ${\cal M}_3$ are depicted by $\square$, $\circ$, $\times$, and $\star$ respectively.}
\label{fig:numexss1}
\end{figure}

\begin{figure}[ht]
\centerline{
\epsfig{file=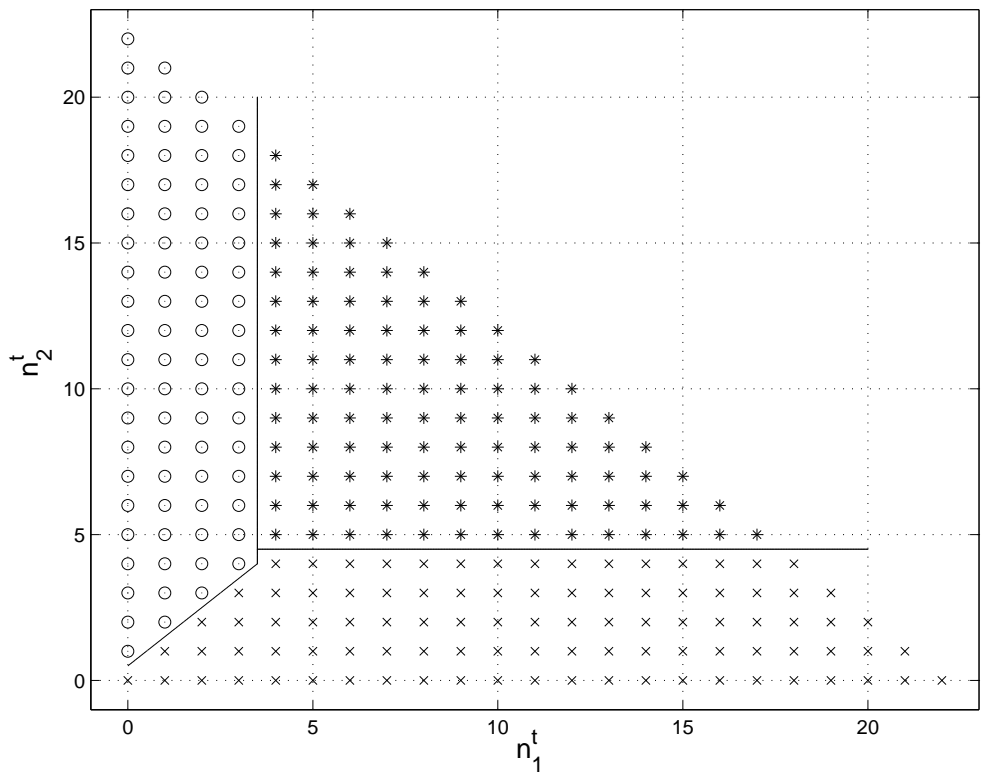,scale=0.75}}
\caption{Partitioning of the $(n_1^t,n_2^t)$ plane into decision regions for fixed $n_3^t=8,T=40,t=30$ for $p=0.3$. The states in which it is optimal to schedule ${\cal M}_0, {\cal M}_1, {\cal M}_2$, and ${\cal M}_3$ are depicted by $\square$, $\circ$, $\times$, and $\star$ respectively.}
\label{fig:numexss2}
\end{figure}

\begin{numex}
Consider a system with three meta-queues ($N=3$) and a time-horizon of $T=40$ time-slots. Let us fix $n_3^t=8$ and look at the projection of the three dimensional state-space on the $(n_1^t,n_2^t)$ plane, for $t=30$. The entries in the TSPs of the meta-queues were generated from an i.i.d. Bernoulli process with parameter $p$. Fig.~\ref {fig:numexss1} and Fig.~\ref {fig:numexss2}  depict the partitioning of the $(n_1,n_2)$ plane for $p=0.1$ and $p=0.3$, respectively. Since the TSPs are more sparse in the case $p=0.1$ (more relaxed deadlines), it is optimal to idle the server in several states.
\end{numex}

\subsection{The myopic/greedy policy}
\label{ssec:myopicss}

The complexity of computing the optimal policy $\widetilde{\Pi}_T^\star$ increases exponentially in both $T$ and $N$. However, the per time-slot complexity of the myopic/greedy policy associated with the problem is only ${\cal O}(N)$. The myopic policy schedules ${\cal M}_{i^\star}$ in the $t^{th}$ time-slot in state $\n$ such that
\begin{equation}
i^\star = \mathop{\arg \min}_{j=0,\ldots,N} \{ \Psi(\n + \e_j - \Xwb^t) \}.
\label{myopicss}
\end{equation}
Once again, the myopic policy for the aforementioned scheduling problem is provably optimal for the case $N=2$. The proof is similar to the proof of Theorem \ref {theorem:2x2myopic} and is omitted.

\subsection{The Largest Lag First (LLF) policy}
\label{ssec:llf-mq}
Consider the special case of quadratic cost functions for concreteness. It is easily seen that the myopic policy in this case selects ${\cal M}_{i^\star}$ such that
\begin{equation}
i^\star = \left\{
\begin{array}{ccc}
\tilde{i}^\star & ; & 2(\dw_{\tilde{i}^\star} - \xw_{\tilde{i}^\star}^t) + 1 < 0 \\
0 & ; & \text{else},
\end{array}
\right.
\label{myopicssquad}
\end{equation}
where
\eq
\displaystyle \tilde{i}^\star \triangleq \mathop{\arg \min}_{j=1,\ldots,N} \{ \dw_j-\xw_j^t \}.
\en
The arguments are similar to those given in Example \ref {numex:qcf}. 

We refer to the policy in (\ref {myopicssquad}) as {\bf Largest Lag First} (LLF), because it chooses the meta-queue with the most negative deviation (equivalently, largest lag). The LLF policy idles the server if the deviations of all meta-queues are non-negative.

\subsection{Meta-queue based service trace control}
\label{ssec:single}

Having studied the salient features of the single server scheduling problem, we revert our attention to the service trace control problem for the IQ switch. Suppose that the STC chooses a configuration from subset ${\cal S}_\v$ alone, and VOQ deviations are mapped to meta-queue deviations through a mapping $\Gamma$. Let ${\cal I}_i=\{i_1,\ldots,i_N\}$ denote the set of VOQs which are served by the $i^{th}$ configuration in ${\cal S}_\v$, namely ${\cal C}^{i-1}(\v)$. These VOQs constitute the $i^{th}$ meta-queue in the single server system. Given deviation vector $\d$ for the switch, we denote the deviation of the $i^{th}$ meta-queue by $\Gamma(\d;{\cal I}_i)$. The LLF policy of Section \ref {ssec:llf-mq} then schedules the $j^{th}$ meta-queue such that
\eq
\displaystyle j = \mathop{\arg \min}_{i=1,\ldots,N} \{ \Gamma(\d;{\cal I}_i) \}.
\en
Once a meta-queue is chosen, a partial configuration is extracted by idling VOQs with non-negative deviation. We now examine two special choices of $\Gamma$.

\subsubsection{The MSL-SS policy revisited}
Consider $\Gamma(\d;{\cal I}) = \displaystyle \sum_{j \in {\cal I}} d_j$; With this choice of $\Gamma$, we recover the MSL-SS policy proposed in Section \ref {ssec:msls}. Thus, the service trace control problem for an IQ switch with single subset operation is equivalent to the single server scheduling problem if all cost functions (both for the VOQs and the meta-queues) are quadratic and the deviation of a meta-queue is defined as the sum of the deviations of its constituent VOQs.

\subsubsection{The LLF-SS policy}
\label{sssec:llfss}
Consider $\displaystyle \Gamma(\d;{\cal I}) = \mathop{\min}_{j \in {\cal I}} \{ d_j \}$; For this $\Gamma$, the STC in effect selects the VOQ with the largest lag. Since every VOQ is associated with a unique configuration in ${\cal S}_\v$, selecting a VOQ immediately identifies a unique switch configuration. We call this policy the {\bf Largest Lag First - Single Subset} (LLF-SS) policy. The per time-slot complexity of LLF-SS is ${\cal O}(N^2)$, since it involves computing the maximum of an unsorted list of $N^2$ numbers. It can be reduced to ${\cal O}(N)$ through parallelization or efficient data structures (for maintaining dynamic lists).

\begin{remarks}
There is a natural interpretation for the above choice of $\Gamma$. Suppose that the switch is operated using only complete configurations and has been set in configuration ${\cal C}^{i-1}(\v)$ $\tau$ times by the end of the $t^{th}$ time-slot. Denote by $S^t_{i_j}$ the $t^{th}$ entry of the TSP of ${\cal Q}_{i_j}$. It follows that the deviation of the ${\cal Q}_{i_j}$ in the $t^{th}$ time-slot is $d^t_{i_j} = \tau - S^t_{i_j}$, since ${\cal C}^{i-1}(\v)$ serves VOQs indexed by ${\cal I}_i = \{i_1,\ldots,i_N\}$. Now, define
\eq
\displaystyle \Sw_i^t \triangleq \mathop{\max}_{j=1,\ldots,N} \{ S^t_{i_j} \}.
\en
If configuration ${\cal C}^{i-1}(\v)$ is chosen at least $\Sw_i^t$ times by the end of the $t^{th}$ time-slot ($\tau \geq S^t$), all $N$ VOQs indexed by set ${\cal I}_i$ have a non-negative deviation (no lag). We therefore let $\Swb_i = (\Sw_i^1,\Sw_i^2,\ldots)$ be the cTSP of the meta-queue generated by grouping VOQs indexed by set ${\cal I}_i$. It follows that the deviation of the $i^{th}$ meta-queue in the $t^{th}$ time-slot is $\displaystyle \dw^t_i = \tau - \mathop{\max}_{j=1,\ldots,N} \{ S^t_{i_j} \}$, i.e., $\displaystyle \dw^t_i = \mathop{\min}_{j=1,\ldots,N} \{ d_{i_j}^t \}$. In words, the deviation of a meta-queue is the {\it minimum} of the deviation of its constituent VOQs.
\end{remarks}

\begin{remarks}
The meta-queue construction provides a general framework for designing service trace control policies under single subset operation. While we have illustrated the idea with two specific examples here, different families of policies with varying performance tradeoffs can be constructed by appropriately selecting the mapping $\Gamma$, as well as the meta-queue selection policy. For instance, we can set 
$\displaystyle \Gamma(\d;{\cal I}) = \sum_{j \in {\cal I}} c_j d_j$,
where $\{c_j\} \geq 0$ are weight parameters chosen to provide differentiated QoS to VOQs.
\end{remarks}

\section{Admissible Region and Subset Selection}
\label{sec:admit}

Consider a traffic stream with target stream profile (TSP) $\s$ and the corresponding cumulative TSP $\S$. The average ``distance'' between consecutive ``1''s in $\s$ can be interpreted as the {\it average packet inter-departure time target} associated with this traffic stream. By definition, $S^t$ is the number of ``1''s in the TSP in the first $t$ time-slots. We assume that the limit
\eq
\lambda = \displaystyle \mathop{\lim}_{t \rightarrow \infty} \frac{S^t}{t}
\en
exists for every traffic stream, and refer to $1/\lambda$ as the average packet inter-departure time (IDT) target for the traffic stream. Going back to Example \ref {numex:periodic}, where we considered periodic traffic with period $\delta$, we see that $\displaystyle S^t = \lfloor t/\delta \rfloor$ and $\lambda = 1/\delta$.

A larger $\lambda$ implies smaller IDT targets on an average, which means the stream requires more service from the switch. Thus, $\lambda$ can be thought of as the {\bf load} imposed by a stream on the switch. Letting $\lambda_i$ denote the load imposed by the stream at the $i^{th}$ VOQ, we define
\eq
\lambdab \triangleq (\lambda_1,\ldots,\lambda_{N^2})
\en
as the {\bf load vector} for the switch. We now consider a special case where the IDT targets for the traffic stream associated with the $i^{th}$ VOQ are {\it geometrically distributed}
\footnote{For a geometrically distributed random variable $X$ with parameter $p$, the probability mass function is given by $\mathbb{P}[X = k] = (1-p)^{k-1}p, \; k \in \mathbb{N}$.} 
with parameter $\lambda_i \in (0,1)$. Equivalently, every entry in the TSP of ${\cal Q}_i$ is an independent identically distributed (i.i.d.) Bernoulli random variable
\footnote{For a Bernoulli random variable X with parameter $p$, the probability mass function is given by $\mathbb{P}[X=0] = 1-p$ and $\mathbb{P}[X=1] = p$.} 
with mean $\lambda_i$. We refer to this scenario as {\bf i.i.d. loading}. Further, we refer to the scenario $\lambda_i = \lambda, \forall \; i$, i.e., $\displaystyle \lambdab = (\lambda/N^2) \one$ as {\bf uniform i.i.d loading}\footnote{The theory developed here can be extended to the case where TSP entries are generated from a Markov modulated Bernoulli process by considering multi-step drifts of the Lyapunov function (see, for example, \cite {modiano}.)}. Using the notation introduced in (\ref {aaabbb}), the i.i.d. assumption implies:
\begin{eqnarray}
\nonumber \mathbb{E}[\x^t] &=& \lambdab \; \forall \; t \\
\nonumber \mathbb{E}[x_i^t x_j^t] &=& \lambda_i \; \text{if } j=i \; \text{and } \mathbb{E}[x_i^tx_j^t] = 0 \; \text{if } j \neq i, \; \forall \; t \\
\mathbb{E}[x_i^t x_j^\tau] &=& 0 \; \forall \; i,j \; \text{if } \tau \neq t.
\label{iidtsp}
\end{eqnarray}
Next, we define the {\bf admissible region} as the set of all load vectors for which some service trace control policy guarantees finite lags to all VOQs, at all times. Alternatively, if the switch is subject to a load vector not contained in the admissible region, the lag of at least one VOQ grows without bound, regardless of the service trace control policy employed. 
The admissible region for an IQ switch is given by
\begin{equation}
\begin{split}
\Lambda \triangleq \{\lambdab : \sum_{j=1}^N \lambda_{(i-1)N+j} < 1, \; \sum_{j=1}^N \lambda_{j(N-1)+i} < 1, \\ i=1,\ldots,N, \; \lambda_i \in (0,1), \; i=1,\ldots,N^2 \}.
\end{split}
\label{lambdaswitch}
\end{equation}
A policy which ensures finite lags for all VOQs for all load vectors $\lambdab \in \Lambda$ is said to be 100\% admissible. More formally,
\begin{definition}
A policy $\Pi$ is 100\% admissible if $\forall \; \lambdab \in \Lambda$, $\displaystyle \mathop{\lim \inf}_{t \rightarrow \infty} \mathbb{E}[d_i^t] > -\infty \; \forall \; i$, where 
the notation $\mathbb{E}^{\Pi}[\cdot]$ implies that  the expectation is computed under policy $\Pi$.
As a special case, a policy $\Pi$ is 100\% admissible under i.i.d. loading if it satisfies the aforementioned property for all i.i.d. load vectors in $\Lambda$.
\end{definition}

\vspace{0.05in}

\begin{theorem}
$\displaystyle \mathop{\lim \inf}_{t \rightarrow \infty} \mathbb{E}[d_i^t] > -\infty \; \forall \; i$ under the MSL($\ell$) policy, for {\em any admissible} i.i.d. load.
\label{theorem:msltheorem}
\end{theorem}

\vspace{0.05in}

\noindent\begin{proof}
See Appendix \ref {ssec:mslproof}.
\end{proof}

\vspace{0.05in}

We are now ready to answer the two questions raised at the end of Section \ref {sec:config} regarding the efficacy of subset based control. 
Our answer to the first question is that by restricting operation to a single subset, not all load vectors in $\Lambda$ can be supported. However, all uniform loads can be supported. In particular,

\vspace{0.05in}

\begin{theorem}
$\displaystyle \mathop{\lim \inf}_{t \rightarrow \infty} \mathbb{E}[d_i^t] > -\infty \; \forall \; i$ under the LLF-SS policy, for any {\it admissible uniform} i.i.d load, independent of the choice of operational subset.
\label{theorem:llfsstheorem}
\end{theorem}

\vspace{0.05in}

\noindent\begin{proof}
See Appendix \ref {ssec:llfssproof}.
\end{proof}

\vspace{0.05in}

 \begin{theorem}
$\displaystyle \mathop{\lim \inf}_{t \rightarrow \infty} \mathbb{E}[d_i^t] > -\infty \; \forall \; i$ under the MSL($\ell$)-SS policy, for any {\it admissible uniform} i.i.d. load, independent of the choice of operational subset.
\label{theorem:mslsstheorem}
\end{theorem}

\vspace{0.05in}

\noindent\begin{proof}
See Appendix \ref {ssec:mslssproof}.
\end{proof}

\vspace{0.05in}

It must be noted that there exists a non-empty subset of non-uniform load vectors in $\Lambda$ (even near the ``boundary'' of $\Lambda$) under which LLF-SS guarantees bounded lags to all VOQs, if the operational subset is suitably chosen.

\begin{numex}
Consider $\lambdab = (1-\epsilon,0,0,0,0,1-\epsilon,0,1-\epsilon,0)$ for a $3 \times 3$ switch, for some $\epsilon \in (0,2/3)$. In this case, operating LLF-SS with the first subset (generated by $\v_1$) in Fig. \ref {fig:ortho} cannot guarantee bounded lags to all VOQs, while operating the same policy with the second subset (generated by $\v_2$) can guarantee bounded lags.
\end{numex}

It is possible to construct non-uniform i.i.d. load vectors in $\Lambda$ under which LLF-SS cannot guarantee bounded lags to all VOQs, irrespective of the choice of the operational subset.

\begin{numex}
Consider $\lambdab = (c,0,0,0,c/2,c/2,0,c/2,c/2)$ where $c = 1-\epsilon$ for some $\epsilon \in (0,1/2)$. The LLF-SS policy cannot guarantee bounded lags to all VOQs, regardless of the choice of operational subset.
\end{numex}
Similar examples can be constructed for the MSL($\ell$)-SS policy.

\begin{table*}
	\centering
		\begin{tabular}{|l|l|c|c|c|}
		 \hline
		 {\bf Policy} & {\bf Brief description} & {\bf Complexity} & {\bf 100\% admissibile} & {\bf Knows $\lambda$}\\
		 \hline
		 MSL & Pick configuration (from ${\cal V}$) with maximum sum of VOQ lags & ${\cal O}(N^3)$ & $\checkmark$ & $\times$\\
		 \hline
		 MSL-SS & Pick configuration (single subset) with max sum of VOQ lags & ${\cal O}(N^2)$ & $\times$ & $\times$\\
		 \hline
		 LLF-SS & Pick configuration (single subset) with most lagged VOQ & ${\cal O}(N^2)$ & $\times$ & $\times$ \\
		 \hline
		 MSL-RS & Randomized subset selection + MSL-SS & ${\cal O}(N^2)$ & $\checkmark$ & $\checkmark$ \\
		 \hline
		 LLF-RS & Randomized subset selection + LLF-SS & ${\cal O}(N^2)$ & $\checkmark$ & $\checkmark$ \\
		 \hline
		 MSL-pSEL(P) & Periodic subset selection (with period $P$) + MSL-SS & ${\cal O}(N^2)$ & $\checkmark$ & $\times$ \\
		 \hline
		 LLF-pSEL(P) & Periodic subset selection (with period $P$) + LLF-SS & ${\cal O}(N^2)$ & $\checkmark$ & $\times$\\
		 \hline
			
		\end{tabular}
		\caption{Key properties of some service trace control policies proposed in the paper.}
		\label{table:policy}
\end{table*}

\subsection{Randomized subset selection}
\label{ssec:randsel}

As we saw in the previous section, service trace control based on single subset operation is not enough to support all admissible loads. However, subset based operation in conjunction with an appropriate subset selection policy can achieve the desired goal. We propose one such subset selection policy in this section. To this end, denote the $k^{th}$ configuration vector by $\v_k$ and the corresponding generated subset by ${\cal S}_k = \{ {\cal C}^i(\v_k)\}_{i=0}^{N-1}$. Consider the Birkoff von Neumann (BV) decomposition \cite{chang1} of load vector $\lambdab \in \Lambda$ given by
\begin{equation}
\lambdab = \sum_{k=1}^{(N-1)!} \sum_{i=0}^{N-1} \zeta_{ik} {\cal C}^i(\v_k), \quad \sum_{k=1}^{(N-1)!} \sum_{i=0}^{N-1} \zeta_{ik}=\zeta<1.
\label{bv}
\end{equation}
Define a probability distribution on the subsets by
\eq
\displaystyle \theta_k \triangleq \frac{1}{\zeta} \sum_{i=0}^{N-1} \zeta_{ik}, \; k = 1,2,\ldots,(N-1)!
\label{thetadef}
\en
Now, consider the following two-step service trace control policy, namely {\bf Maximum Sum of Lags - Random Subset } (MSL-RS), which combines the MSL-SS policy of Section \ref {ssec:msls} with the notion of randomized subset selection:
\begin{enumerate}
\item Select configuration subset ${\cal S}_k$ with probability $\theta_k$.
\item Select a configuration from ${\cal S}_k$ based on MSL-SS.
\end{enumerate}
The computational complexity of MSL-RS is ${\cal O}(N^2)$ per time-slot, since MSL-SS has complexity ${\cal O}(N^2)$ and the BV decomposition of $\lambdab$ contains at most $N^2-2N+2$ non-zero terms \cite {chang1}. 
 
\vspace{0.05in}

\begin{theorem}
$\displaystyle \mathop{\lim \inf}_{t \rightarrow \infty} \mathbb{E}[d_i^t] > -\infty \; \forall \; i$ under MSL($\ell$)-RS, for {\it any admissible} i.i.d. load.
\label{theorem:mslrstheorem}
\end{theorem}

\vspace{0.05in}

\noindent\begin{proof}
See Appendix \ref {ssec:mslrsproof}.
\end{proof}

\vspace{0.05in}

The {\bf Largest Lag First - Random Subset} (LLF-RS) policy is constructed analogously, by combining the idea of randomized subset selection with the LLF-SS policy of Section \ref {sssec:llfss}. Finally, note that the MSL-RS and LLF-RS policies can be extended to construct the MSL($\ell$)-RS and LLF($\ell$)-RS families of policies, respectively (discussed in Remark \ref {remarks:mslL}). These policies allow traffic streams to enjoy a lead of up to $\ell > 0$, instead of idling them when they are not lagging.

\subsection{Periodic subset selection}
\label{sec:heuristic}

\subsubsection{The pSEL(P) rule}
The MSL($\ell$)-RS policy proposed in the previous section can support all admissible loads {\it and} has low computational complexity. However, the policy requires a priori knowledge of the load vector $\lambdab$. We, on the other hand, are interested in designing robust control policies which do not rely on statistical knowledge about the input traffic streams. Thus, to eliminate dependence on $\lambdab$, we propose the following {\it periodic} subset selection rule: Suppose the switch is currently being operated using configuration subset ${\cal S}_\v$. Every $P > 0$ time-slots, a complete configuration $\v^\star$ is selected, based on some service trace control policy. If $\v^\star \in {\cal S}_\v$, the switch continues to operate with configuration subset ${\cal S}_\v$, otherwise the switch starts operating in the configuration subset generated by $\v^\star$, viz., ${\cal S}_{\v^\star} = \{\v^\star, {\cal C}(\v^\star),\ldots,{\cal C}^{N-1}(\v^\star)\}$. Once a configuration subset has been selected, the switch can be operated using any subset based service trace control policy (e.g. MSL-SS). We refer to this subset selection rule as {\bf pSEL(P)} ({\bf P}eriodic {\bf Sel}ection with period $P$).

\subsubsection{The MSL-pSEL(P) policy}
\label{ssec:mslpselp}
We combine the pSEL(P) subset selection rule with the MSL policy of Section \ref {ssec:msl} and the MSL-SS policy of Section \ref {ssec:msls} to propose the {\bf Maximum Sum of Lags - Periodic Selection (P)} (MSL-pSEL(P)) service trace control policy. Every $P$ time-slots, the MSL-pSEL(P) policy computes  the switch configuration $\v^\star$ based on the MSL policy. If $\v^\star$ is in the current operational subset, the MSEL-pSEL(P) policy continues to operate the switch using the current subset, otherwise it switches to the configuration subset generated by $\v^\star$, viz., $\{\v^\star,{\cal C}^(\v^\star),\ldots,{\cal C}^{N-1}(\v^\star)\}$. 
In the intermediate $P-1$ time-slots, MSL-pSEL(P) operates the switch using the MSL-SS policy. 

The per time-slot complexity of the MSL policy is ${\cal O}(N^3)$. This computation needs to be done every $P$ time-slots to update the configuration subset. The complexity of the MSL-SS policy is ${\cal O}(N^2)$, as discussed in \ref {ssec:msls}. The MSL-SS policy needs to be executed in the $P-1$ intermediate time-slots between configuration subset updates. Thus, the computational complexity of MSL-pSEL(P) is ${\cal O}\left( {N^3}/{P} + (1 - {1}/{P} ) N^2 \right)$. If $P = {\cal O}(N)$, i.e., if the configuration subset is updated roughly every $N$ time-slots for a switch of size $N \times N$, 
the complexity of MSL-pSEL(P) is ${\cal O}(N^2)$.

MSL-pSEL(P) has all the desired traits - a manageable complexity of ${\cal O}(N^2)$, no dependence on load vector $\lambdab$, and as Theorem \ref {theorem:pseltheorem} tells us, it is 100\% admissible under i.i.d. loading.

\vspace{0.05in}

\begin{theorem}
$\displaystyle \mathop{\lim \inf}_{t \rightarrow \infty} \mathbb{E}[d_i^t] > -\infty \; \forall \; i$ under MSL-pSEL(P), for {\it any admissible} i.i.d. load.
\label{theorem:pseltheorem}
\end{theorem}

\vspace{0.05in}

\noindent\begin{proof}
See Appendix \ref {ssec:pselproof}.
\end{proof}

\vspace{0.05in}

We close this section by noting that 
the {\bf Largest Lag First - Periodic Selection (P)} (LLF-pSEL(P)) policy is constructed by combining the LLF policy (Section \ref {ssec:llf-mq}) and the LLF-SS policy (Section \ref {sssec:llfss}) with the pSEL(P) rule.

\section{Performance Evaluation}
\label{sec:simulation}

In this section, we present simulation results to characterize the performance of the proposed service trace control policies. 

\subsection{Simulation setup}
All results presented here are for a $16 \times 16$ IQ switch. We contrast the performance of the following policies: 
\begin{itemize}
\item {\it Maximum Sum of Lags (MSL)}: This policy was proposed in Section \ref {ssec:msl}. MSL computes the maximum weight matching (with VOQ lags as edge weights) over all possible switch configurations (set of size $N!$), and will be the benchmark for all other lower complexity policies.
\item {\it Maximum Sum of Lags - Single Subset (MSL-SS)}: This policy was proposed in Section \ref {ssec:msls}. MSL-SS computes the maximum weight matching over one configuration subset only (using VOQ lags as edge weights).
\item {\it Largest Lag First - Single Subset (LLF-SS)}: This policy was proposed in Section \ref {sssec:llfss}. LLF-SS operates on a single configuration subset, and picks the VOQ (and hence the configuration, since each VOQ is associated with a unique configuration within a subset) with the largest lag.
\item {\it Maximum Sum of Lags - Periodic Selection (MSL-pSEL(16))}: This policy was proposed in Section \ref {ssec:mslpselp}. The behavior is similar to MSL-SS, except that the underlying operational configuration subset is updated every 16 slots.
\item {\it Largest Lag First - Periodic Selection (MSL-pSEL(16))}: This policy was proposed in Section \ref {ssec:mslpselp}. The behavior is similar to LLF-SS, except that the underlying operational configuration subset is updated every 16 slots.
\end{itemize}
Salient features of the above policies are enumerated in Table \ref {table:policy}. All policies require a two-step implementation. A complete switch configuration is selected in the first step and all VOQs with non-negative deviations are idled to extract a partial configuration in the second step. Thus, no VOQ can lead under any of the policies under consideration. We also simulated the performance of policies which allow VOQs to acquire a lead of up to $\ell>0$, but did not observe any relative difference in the performance of different policies for fixed $\ell$. 

We consider the following two performance metrics:
\begin{itemize}
\item {\it Average Deviation}: Empirical mean of VOQ deviations, averaged over all $16^2=256$ VOQs.
\item {\it Variance}: Empirical variance of VOQ deviations, averaged over all 256 VOQs.
\end{itemize}
Since no VOQ can lead in our simulation setup, the deviation of every VOQ (and hence the average deviation for every policy) is upper bounded by zero.

Clearly, we want both the mean and variance of the deviations to be close to zero, for all traffic streams. A mean close to zero with large variance is not sufficient, since it indicates severe instantaneous positive/negative distortion of the target profiles. In other words, it implies that the output of the switch is not ``smooth''. This is not ideal from a flow control perspective, since a bursty output stream from the switch makes scheduling and buffering at downstream switches harder. Also, a small variance with a large non-zero mean is undesirable, since it indicates that one or more VOQs are missing their deadlines frequently.
\begin{remarks}
Ideally, all policies should be benchmarked relative to the optimal service trace control policy, which is computed by solving the DP equations in (\ref {dp1}). However, the complexity of evaluating the optimal policy grows exponentially with the size of the switch, viz. $N$ and the length of the time horizon of interest, viz. $T$. In our setup, $N=16$ and $T=50,000$. The complexity of solving the DP equations for a problem of this magnitude is simply prohibitive. We therefore resort to the next best option, i.e., using the myopic policy (MSL) as a performance benchmark. Recall that we have analytically proven the optimality of the myopic policy for $N=2$ (see Theorem \ref {theorem:2x2myopic}) and numerically verified that it is ``close'' to being optimal for $N=3,4$. Note that even the myopic policy is quite expensive to implement, since it entails computing a maximum weight matching in every time-slot.
\end{remarks}

\subsection{Discussion of simulation results}
We report simulation results for four distinct loading scenarios. Every point on the performance curves depicted here was generated by averaging over 50,000 time-slots. 

\begin{figure}[ht]
\centerline{
\epsfig{file=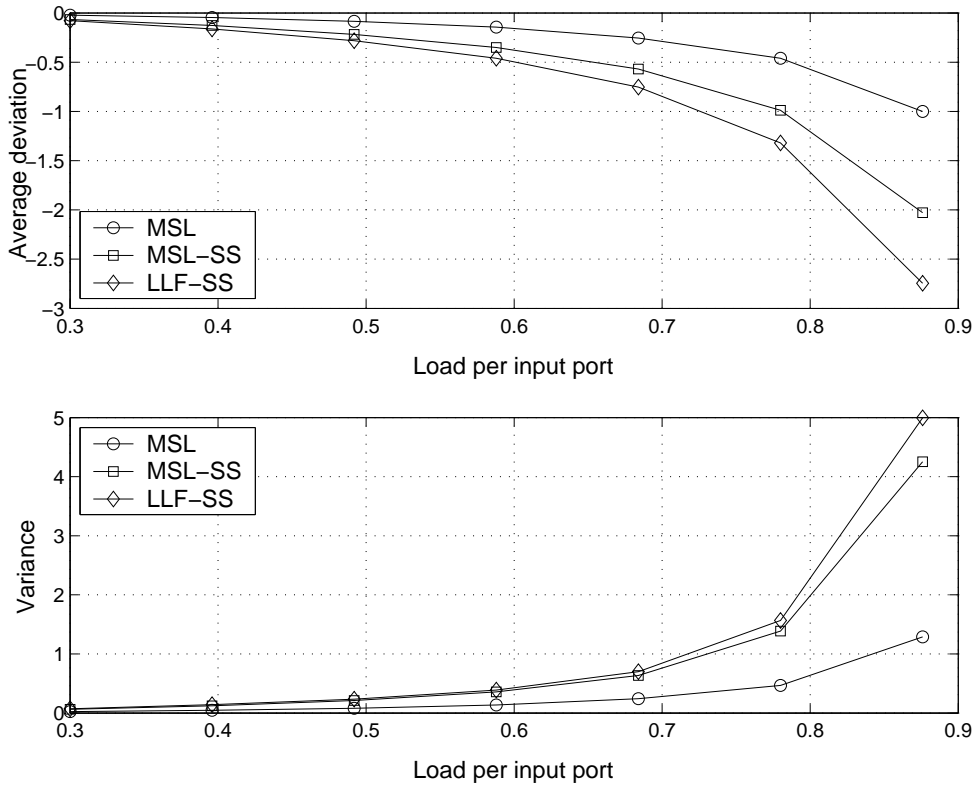,scale=0.75}}
\caption{Performance of MSL, MSL-SS and LLF-SS under uniform i.i.d. loading for an $16 \times 16$ switch}
\label{fig:geomunif}
\end{figure}

\begin{figure}[ht]
\centerline{
\epsfig{file=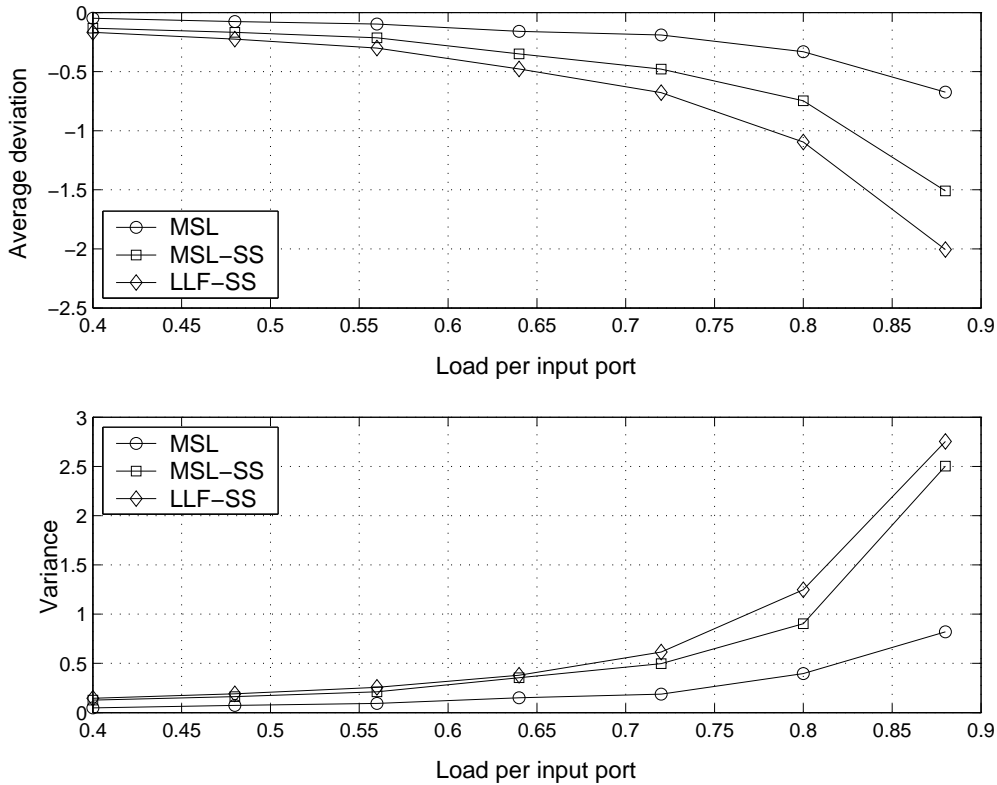,scale=0.75}}
\caption{Performance of MSL, MSL-SS and LLF-SS under parallel-heavy i.i.d. loading for an $16 \times 16$ switch}
\label{fig:geomnonunif1}
\end{figure}

\begin{figure}[ht]
\centerline{
\epsfig{file=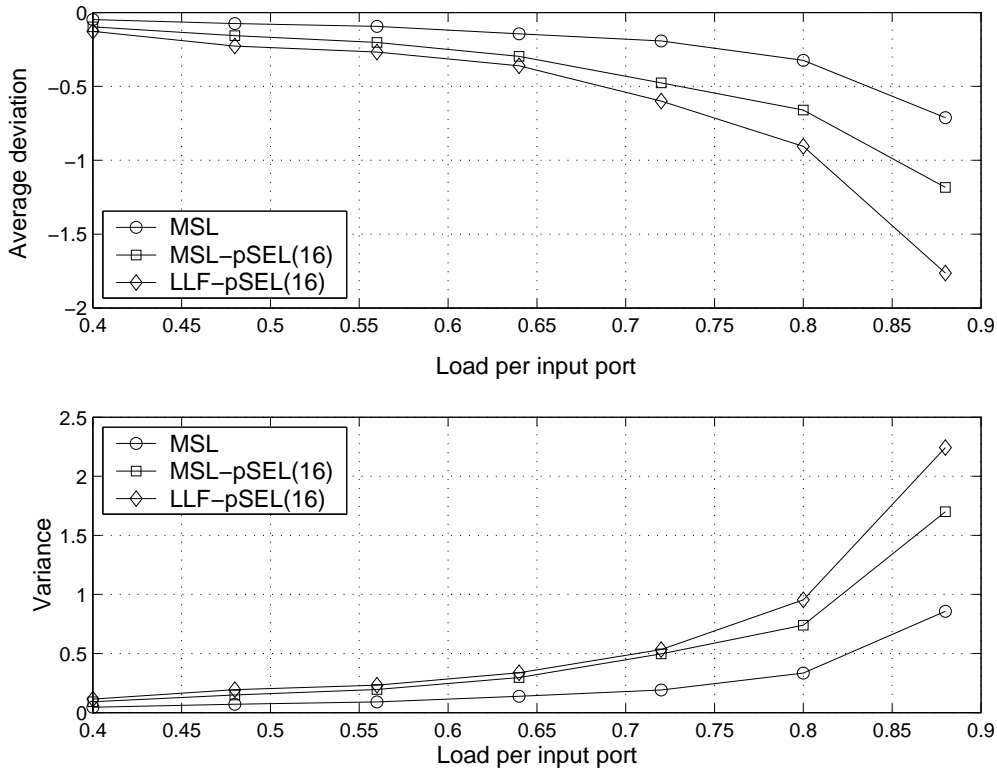,scale=0.75}}
\caption{Performance of MSL, MSL-pSEL(16) and LLF-pSEL(16) under cross-heavy i.i.d. loading for an $16 \times 16$ switch}
\label{fig:geomnonunif2}
\end{figure}

\begin{figure}[ht]
\centerline{
\epsfig{file=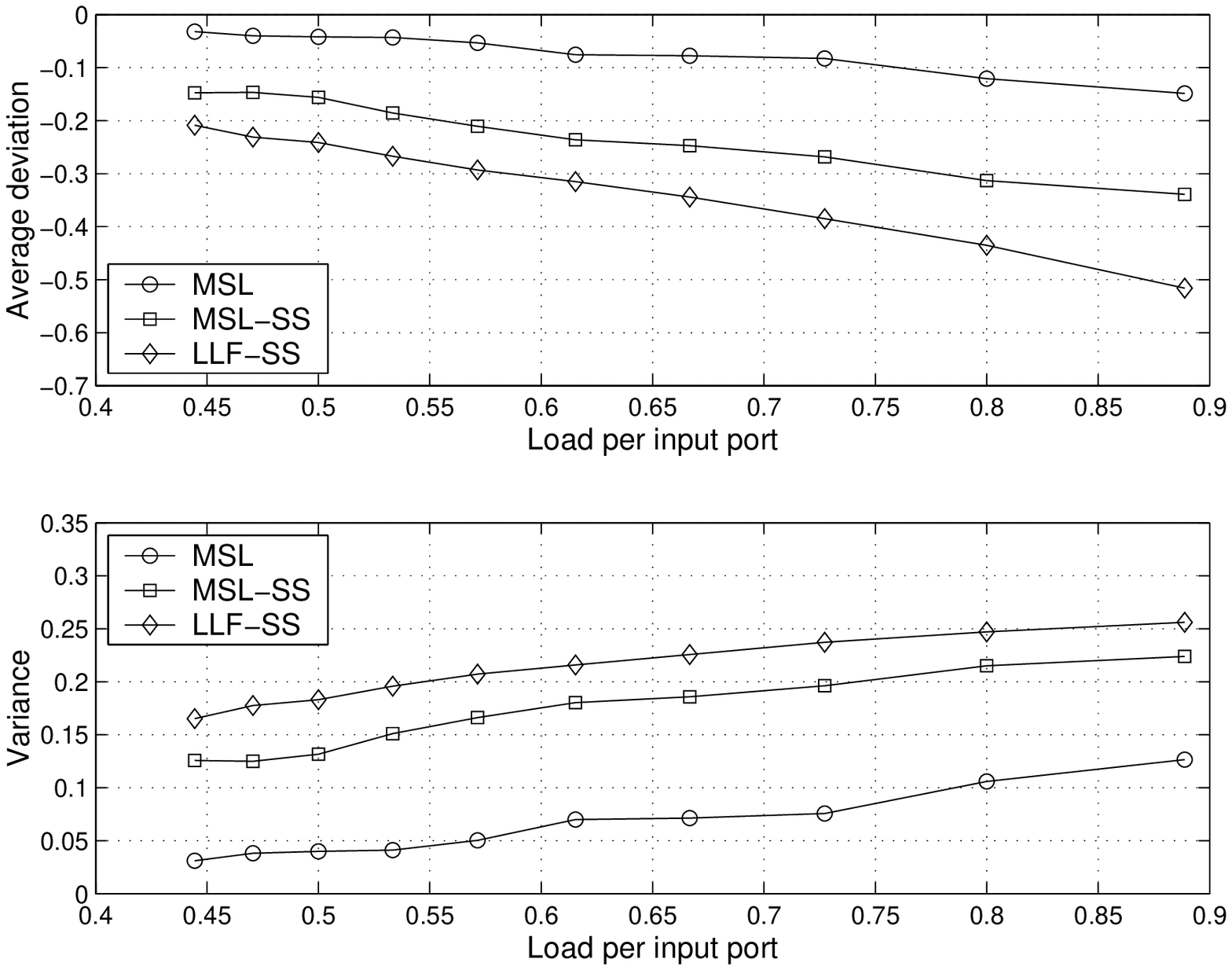,scale=0.75}}
\caption{Performance of MSL, MSL-SS and LLF-SS under uniform periodic loading for an $16 \times 16$ switch}
\label{fig:periodunif}
\end{figure}

\subsubsection{Uniform i.i.d. loading}
In this case, all streams have geometrically distributed inter-departure time targets with parameter $\lambda/16$. The load per input port is therefore $16 \times \lambda/16 = \lambda$ (number of VOQs $\times$ load per stream/VOQ). The performance of the MSL, MSL-SS, and LLF-SS policies is depicted in Fig. \ref {fig:geomunif}. The results show that operating the switch with a single subset works quite well if the switch is uniformly loaded, especially up to $\sim 60\%$ loading of the switch. The loss in performance vis-a-vis the MSL policy comes with a significant reduction in complexity. Recall that the MSL policy has to perform a full-blown maximum weight matching computation in every time-slot. Subset selection is not needed in the uniform loading scenario (Theorems \ref {theorem:llfsstheorem} and \ref {theorem:mslsstheorem}). It suffices to operate the switch using only one configuration subset, which can be chosen arbitrarily.

\subsubsection{Parallel-heavy i.i.d loading}
In this case, all traffic streams associated with ``parallel'' VOQs have geometrically distributed inter-departure time targets with parameter $\lambda_1$ and ``diagonal'' VOQs have geometrically distributed inter-departure time targets with parameter $\lambda_2 < \lambda_1$. We call a VOQ parallel if it buffers packets destined from the $i^{th}$ input port to the $i^{th}$ output port for $i \in \{1,\ldots,16\}$, and call it diagonal otherwise. Thus, a $16 \times 16$ switch has 16 parallel VOQs and $16^2-16 = 240$ diagonal VOQs. We fixed $\lambda_2$ and varied $\lambda_1$ to vary the load per input port, viz. $\lambda_1 + 15\lambda_2$. Performance results are depicted in Fig. \ref {fig:geomnonunif1}. For MSL-SS and LLF-SS, we selected the subset generated by the configuration which concurrently serves all 16 parallel VOQs. Since this configuration needs to be selected frequently, especially as $\lambda_1$ increases, single subset operation based policies perform quite well in this non-uniform loading scenario. Note that subset based policies would perform poorly in this scenario if the configuration subset is not selected appropriately. Good performance can however be achieved by combining subset based operation with the periodic subset selection rule, as illustrated by the next set of simulation results.

\subsubsection{Cross-heavy i.i.d. loading}
Once again, all traffic streams have geometrically distributed inter-departure time targets. However, VOQs which buffer packets destined from input port $i$ to output port $i+1$ (for odd $i$) and from input port $i$ to output port $i-1$ (for even $i$) are more heavily loaded (parameter $\lambda_1$) than other VOQs (parameter $\lambda_2<\lambda_1$). We allude to this scenario as cross-heavy loading because the configuration which serves the more heavily loaded VOQs forms a criss-cross pattern with eight ``crosses'' ($\times$). We fixed $\lambda_2$ and varied $\lambda_1$ to vary the load per input port, viz. $\lambda_1+15\lambda_2$. For MSL-SS and LLF-SS, we used the same configuration subset as for the parallel-heavy i.i.d. loading experiment. Since the cross pattern is not contained in this subset, the performance of single subset based policies degrades severely as $\lambda_1$ increases, and is therefore not depicted here. However, periodic subset selection, in conjunction with single subset operation delivers performance quite close to the benchmark MSL policy, especially for switch loading up to $\sim 65\%$. Recall that this performance is achieved at a much lower computational complexity. The results are depicted in Fig. \ref {fig:geomnonunif2}. 
The MSL-SS and LLF-SS policies would have performed well if we had chosen the configuration subset which serves the most heavily loaded VOQs (the ones forming the cross pattern) as the operational subset. This is exactly what we had done in the diagonal loading scenario. However, it is not always possible for the switch controller to have a priori information about the loading pattern. Subset based operation therefore needs to be combined with a subset selection rule to ensure good performance in unknown loading scenarios.

\subsubsection{Uniform periodic loading}
In this case, all traffic streams have identical inter-departure time targets equal to $\delta$ (see Example \ref {numex:periodic}). The streams are offset relative to each other, i.e., the TSPs of all streams are time-shifted versions of each other. For instance, suppose the desired departure times of packets from one of the stream are $(\tau,\tau+\delta,\tau+2\delta,\ldots)$ for some $\tau \in \mathbb{Z}_+,\delta \in \mathbb{N}$. The desired departure times of packets from another stream traversing the same switch could be $(\tau+\tau',\tau+\tau'+\delta,\tau+\tau'+2\delta,\ldots)$ for some $\tau' \in \mathbb{Z}_+$. Both streams are periodic with IDTs equal to $\delta$, but are offset by $\tau'$ slots with respect to each other. We generated the offsets uniformly at random from the set $\{0,1,\ldots,\delta-1\}$ and varied $\delta$ from 18 to 36 slots to vary the load per input port, viz. $16/\delta$. The performance results are depicted in Fig. \ref {fig:periodunif}. 
The efficacy of single subset based operation under uniform switch loading is evident from the plots. For instance, even at $\sim 80\%$ loading of the switch, the average deviation and the variance of the deviation under the MSL-SS policy are -0.3 and 0.2, respectively. This means that on an average, the received service traces for all 256 VOQs deviate from the target stream profiles by only 0.3 time-slots, with a standard deviation of $\approx 0.45$ time-slots. Such small negative deviations with small variances can easily be corrected for by allowing traffic streams to gain a lead of 1-2 time-slots (e.g. by using the MSL($\ell$)-SS policy). Roughly speaking, using MSL($\ell$)-SS adds $\ell$ time-slots to the average deviation, without impacting the variance. For $\ell = 2$, the average deviation for the specific example discussed above would be 1.7. With a standard deviation of 0.44, the probability of missing a deadline (target departure time) will be minimal, even at $\sim 80\%$ loading of the switch. 

We also evaluated the performance of the proposed policies under non-uniform periodic loading and Markovian modulated Bernoulli loading (entries of the target stream profile were generated from an MMB process, instead of an i.i.d. Bernoulli process). 
The performance was observed to be more or less invariant to the statistics of the input traffic streams, underlining the robustness of the proposed policies.

\begin{remarks}
Our simulation results demonstrate that it is possible to render IQ switches nearly {\it transparent} to deadline sensitive traffic streams by minimizing the distortion of their target profiles. Moreover, this can be accomplished with low complexity online scheduling policies, and with no prior knowledge of the input traffic statistics. For the proposed policies, the transparency of the switch is particularly strong under moderate loading ($<60\%$), which is a very relevant regime, since a switch is unlikely to be loaded to capacity with real-time traffic. For example, at $\sim 50\%$ loading, for all four loading scenarios simulated here, the average deviation from the target stream profile under all proposed policies is no more -0.3, with a variance below 0.2. Moreover, the proposed policies do not require any prior knowledge of the statistics of the input traffic. For instance, they do not require the streams to be periodic or to be constrained by a leaky bucket, as long as the offered load is within the admissible region of the switch. 
Finally, the ${\cal O}(N^2)$ complexity of the proposed policies render them more amenable to implementation in high performance packet switches vis-\`a-vis other schedulers proposed in the literature for multiclass periodic traffic, which have a computational complexity of ${\cal O}(N^3)$ or ${\cal O}(N^4)$, in addition to their high implementation complexity.
\end{remarks}

\section{Conclusion}
\label{sec:conclude}

We examined the problem of packet switch scheduling for minimizing aggregate distortion of outflow traffic streams with respect to target packet inter-departure times. The study was initially motivated by the need to provide QoS for real-time multimedia traffic over packet networks. The notion of switch configuration subset based control was leveraged to design robust, low complexity, near optimal schedules amenable to implementation in high performance packet switches. Such schedules have been shown to achieve 100\% pull-throughput under certain natural statistics of target profiles.

Many theoretical questions remain open, including the pull-throughput region of the switch under general target profile statistics. Moreover, sweeping experimentation is needed to scope out the design and performance of such switches and schedules in broad, diverse target profile regimes.

\section{Appendix}
\label{sec:appendix}

\subsection{Proof of Theorem \ref {theorem:2x2myopic}}
\label{ssec:proof1}

\noindent\begin{proof}
For ease of exposition, we only treat the case where the switch is never idled. A $2 \times 2$ switch can be set into two possible configurations, $\v_1=(1 \; 0 \; 0 \; 1)$ and $\v_2= (0 \; 1 \; 1 \; 0)$. Given initial state $\d^0$, we say that state $\d$ is {\it reachable} in the $t^{th}$ time-slot if there is a sequence of configurations which drive the switch to state $\d$ in $t$ time-slots. The reachable states in the $t^{th}$ time-slot constitute the set ${\cal R}^t = \{\d_k^t \triangleq \d^0 + k\v_1 + (t-k)\v_2 - \X^t\}_{k=0}^t$. The state $\d_k^t$ is reached if the STC selects configuration $\v_1$ in $k$ of the first $t$ time-slots and $\v_2$ in remaining $t-k$ time-slots. The states reachable in the $(t+1)^{st}$ time-slot given state $\d_k^t$ in the $t^{th}$ time-slot are $\d_k^t +\v_1 - \x^{t+1} = \d_{k+1}^{t+1}$ and $\d_k^t + \v_2 - \x^{t+1} = \d_{k}^{t+1}$. Equivalently, we can identify the state in the $t^{th}$ time-slot by the index $k$, which increases by 1 in the next time-slot if $\v_1$ is chosen and remains unaltered if $\v_2$ is chosen.

Observe that $\d_k^t$ is a sum of two components:
\begin{enumerate}
\item $\d^0 + k\v_1 + (t-k)\v_2$, the evolution of which is determined by the service trace control policy
\item $-\X^t$, the evolution of which is determined by the inter-departure time targets of the input streams. 
\end{enumerate}
The first component can be represented using a {\it directed acyclic graph} (DAG). The root of the DAG is at $\d^0$. Nodes of the DAG at depth $t$ correspond to the policy dependent component of the reachable states in the $t^{th}$ time-slot. There are $t+1$ nodes at depth $t$, ordered in increasing order of index $k$ from right to left (Fig. \ref {fig:tree}). The optimal policy traverses the least cost path from the root to one of the leaves.

We use $C(\d_k^t)=i$ to denote that $\Pi_T^\star$ chooses configuration $\v_i$ in the $t^{th}$ time-slot in state corresponding to index $k$. Also, we define 
\begin{equation}
\Omega^t(\d) \triangleq V^t(\d) + \Phi(\d).
\end{equation}
\begin{equation}
\gamma^t(\d) \triangleq \Omega^{t+1}(\d+\v_1-\x^t) -\Omega^{t+1}(\d+\v_2-\x^t).
\end{equation}
The quantity $\gamma^t(\d)$ is interpreted as the {\it decision function} in state $\d$ in the $t^{th}$ time-slot, i.e., $C(\d^t) = 1$ (configuration $\v_1$ selected) if $\gamma^t(\d) \leq 0$, and $C(\d^t) = 2$ (configuration $\v_2$ selected) otherwise.

We need four auxiliary results to prove the optimality of the myopic policy. The proofs of the auxiliary results are omitted due to space constraints.
\subsubsection{Auxiliary result 1 (AR1)}
For any state $\d$, $\Phi(\d+\v_1-\v_2)-\Phi(\d) \geq \Phi(\d)-\Phi(\d-\v_1+\v_2)$.

{\em Proof}: Recall from Section \ref {ssec:dpform} that the cost functions $\phi_i(\cdot)$ associated with the VOQs are convex. Thus, for the $i^{th}$ VOQ with deviation $d_i$, the following holds - $\phi_i(d_i+1)-\phi_i(d_i) \geq \phi_i(d_i)-\phi_i(d_i-1)$. Further, from \ref {Phidef}, the cost function $\Phi$ is the sum of cost functions of all VOQs. Combining the convexity of $\phi_i$ with the definition of $\Phi$, we arrive at the desired result.

\subsubsection{Auxiliary result 2 (AR2)}
For $t=1,\ldots,T$ and any state $\d$, $V^{t}(\d+\v_1-\v_2) - V^t(\d) \geq V^t(\d) - V^{t}(\d-\v_1+\v_2)$.

{\em Proof}: The proof is based on inductive arguments. 
{\em Base case}: For the base case, $t=T$, it follows from (\ref {dp1}) and the boundary conditions $V^{T+1}(\d) = 0$ that
\eq
V^T(\d) = \mathop{\min} \left\{ \Phi(\d + \v_1 - \x^T), \Phi(\d + \v_1 - \x^T)	 \right\}.
\en
Suppose the result is true for some $t < T$, i.e.,
\eq
V^{t}(\d'+\v_1-\v_2) - V^t(\d') \geq V^t(\d') - V^{t}(\d'-\v_1+\v_2) \; \forall \; \d'.
\label{temp1}
\en
We will show that the (\ref {temp1}) implies that the result is true for $t+1$, i.e.,
\eq
V^{t+1}(\d'+\v_1-\v_2) - V^{t+1}(\d') \geq V^{t+1}(\d') - V^{t+1}(\d'-\v_1+\v_2) \; \forall \; \d'.
\label{result}
\en
Setting $\d' = \d+\v_1-\x^t$ in (\ref {temp1}) and invoking AR1, we get
\eq
\Omega^{t+1}(\d+2\v_1-\v_2-\x^t) - \Omega^{t+1}(\d+\v_1-\x^t) \geq \Omega^{t+1}(\d+\v_1-\x^t) - \Omega^{t+1}(\d+\v_2-\x^t).
\label{temp2}
\en
Similarly, setting $\d' = \d+\v_2-\x^t$ in (\ref {temp1}) and invoking AR1, we get
\eq
\Omega^{t+1}(\d+\v_1-\x^t) - \Omega^{t+1}(\d+\v_2-\x^t) \geq \Omega^{t+1}(\d+\v_2-\x^t) - \Omega^{t+1}(\d+\v_1-\x^t).
\label{temp3}
\en
It follows from the definition of $\gamma^t(\cdot)$, (\ref {temp2}), and (\ref {temp3}) that
\eq
\gamma^t(\d+\v_1-\v_2) \geq \gamma^t(\d) \geq \gamma^t(\d-\v_1+\v_2).
\label{temp4}
\en

{
\psfrag{ik}{Index $k$}
\psfrag{t}{Time}
\psfrag{r}{Root}
\psfrag{ik2}{Index $k$ increments}
\psfrag{c0}{by $1$ if $\v_1$ is chosen}
\psfrag{c02}{ }
\psfrag{ik3}{Index $k$ remains}
\psfrag{c1}{unaltered if $\v_2$}
\psfrag{c12}{is chosen}
\begin{figure*}[t]
\centerline{\epsfig{file=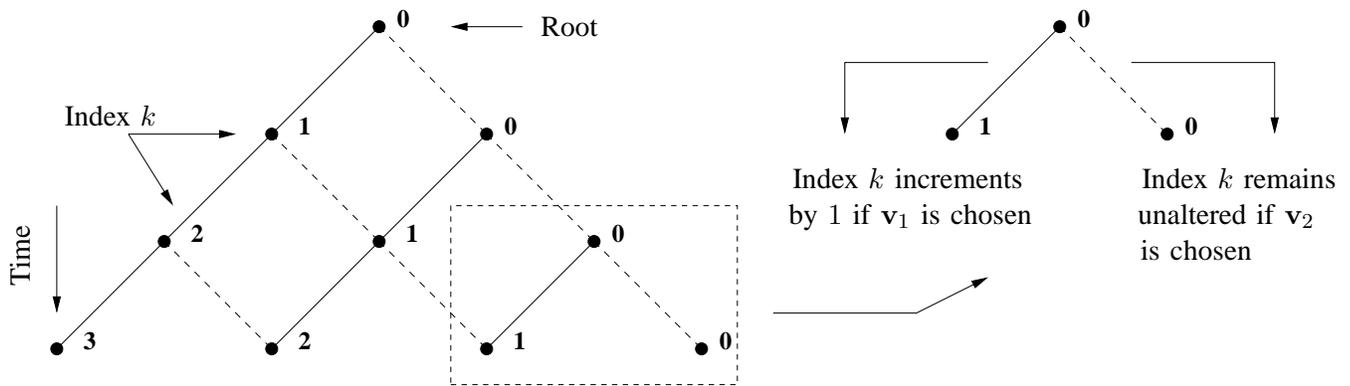,scale=0.75}}
\caption{A DAG based representation of the reachable states in successive time-slots, for a $2 \times 2$ switch. The solid lines represent choice of configuration $\v_1$, while the dotted lines represent choice of configuration $\v_2$. The inset depicts the evolution of state index $k$, based on the choice of configuration in a state.}
\label{fig:tree}
\end{figure*}
}

In view of (\ref {temp4}), four distinct cases arise:
\begin{itemize}
\item $0 \geq \gamma^t(\d+\v_1-\v_2) \geq \gamma^t(\d) \geq \gamma^t(\d-\v_1+\v_2)$: In this case, $C^t(\d-\v_1+\v_2) = C^t(\s) = C^t(\d+\v_1-\v_2) = 1$. We have,
\begin{eqnarray*}
V^t(\d+\v_1-\v_2) &=& \Omega^{t+1}(\d+2\v_1-\v_2-x^t) \\
V^t(\d) &=& \Omega^{t+1}(\d+\v_1-\x^t) \\
V^t(\d-\v_1+\v_2) &=& \Omega^{t+1}(\d+\v_2-\x^t)
\end{eqnarray*}
The result in (\ref {result}) now follows from the (\ref {temp2}) and the above set of equalities.
\item $\gamma^t(\d+\v_1-\v_2) \geq \gamma^t(\d) \geq \gamma^t(\d-\v_1+\v_2) \geq 0$: In this case, $C^t(\d-\v_1+\v_2) = C^t(\s) = C^t(\d+\v_1-\v_2) = 2$. We have,
\begin{eqnarray*}
V^t(\d+\v_1-\v_2) &=& \Omega^{t+1}(\d+\v_1-x^t) \\
V^t(\d) &=& \Omega^{t+1}(\d+\v_2-\x^t) \\
V^t(\d-\v_1+\v_2) &=& \Omega^{t+1}(\d-\v_1+2\v_2-\x^t)
\end{eqnarray*}
The result in (\ref {result}) now follows from the (\ref {temp3}) and the above set of equalities.
\item $\gamma^t(\d+\v_1-\v_2) \geq 0 \geq \gamma^t(\d) \geq \gamma^t(\d-\v_1+\v_2)$: In this case, $C^t(\d-\v_1+\v_2) = C^t(\s) = 1$ and $C^t(\d+\v_1-\v_2) = 2$. We have,
\begin{eqnarray*}
V^t(\d+\v_1-\v_2) &=& \Omega^{t+1}(\d+\v_1-x^t) \\
V^t(\d) &=& \Omega^{t+1}(\d+\v_1-\x^t) \\
V^t(\d-\v_1+\v_2) &=& \Omega^{t+1}(\d+\v_2-\x^t)
\end{eqnarray*}
The result in (\ref {result}) now follows from the above set of equalities, the definition of $\gamma^t(\cdot)$, and the assumption that $\gamma^t(\d) \leq 0$.
\item $\gamma^t(\d+\v_1-\v_2) \geq \gamma^t(\d) \geq 0 \geq \gamma^t(\d-\v_1+\v_2)$: In this case, $C^t(\d-\v_1+\v_2) = 1$ and $C^t(\s) = C^t(\d+\v_1-\v_2) = 2$. We have,
\begin{eqnarray*}
V^t(\d+\v_1-\v_2) &=& \Omega^{t+1}(\d+\v_1-x^t) \\
V^t(\d) &=& \Omega^{t+1}(\d+\v_2-\x^t) \\
V^t(\d-\v_1+\v_2) &=& \Omega^{t+1}(\d+\v_2-\x^t)
\end{eqnarray*}
The result in (\ref {result}) now follows from the above set of equalities, the definition of $\gamma^t(\cdot)$, and the assumption that $\gamma^t(\d) > 0$.
\end{itemize} 
Since the four cases considered above are mutually exhaustive, the proof is complete.

\subsubsection{Auxiliary result 3 (AR3)}
For $t=1,\ldots,T$, $\exists \; k_t^\star \in \{0,\ldots,t\}$ such that $C(\d_k^t)=1 \; \forall \; k \leq k_t^\star$ and $C(\d_k^t)=2 \; \forall \; k > k_t^\star$.

{\em Proof}: Adding the results of AR1 and AR2, and invoking the definition of $\Omega(\cdot)$, we get 
\eq
\Omega^t(\d+\v_1-\v_2) - \Omega^t(\d) \geq \Omega^t(\d) - \Omega^t(\d-\v_1+\v_2). 
\label{temp5}
\en
Combining (\ref {temp5}) with the definition of $\gamma(\cdot)$, it follows that $\gamma^t(\d+\v_1-\v_2) \geq \gamma^t(\d)$. Finally, from the definition of $\d_k^t$, we get $\d_{k+1}^t - d_k^t = \v_1-\v_2$. This implies that $\gamma^t(\d_{k+1}^t) \geq \gamma^t(\d_k^t)$, i.e., the decision function $\gamma^t(\d_k^t)$ is a non-decreasing function of $k$. The proof is based on inductive arguments. We skip the details here. Now, recall that the optimal configuration in state $\d$ at time $t$ is completely determined by the sign of $\gamma^t(\d)$. For fixed $t$, $\gamma^t(\d_k^t)$ can change sign at most once as $k$ increases from 0 to $t$ (by virtue of its monotonicity). In other words, $\exists \; k_t^\star \in \{0,1,\ldots,t\}$ such that $\gamma^t(\d_k^t) \leq 0 \; \forall \; k \leq k_t^\star$ (implying $C(\d_k^t) = 1$) and $\gamma^t(\d_k^t) > 0 \; \forall \; k > k_t^\star$ (implying $C(\d_k^t) = 2$).

\subsubsection{Auxiliary result 4 (AR4)}
$\displaystyle \mathop{\arg \min}_{k=0,\ldots,t} \Omega^t(\d_k^t) = k_t^\star$.

{\em Proof}: We first show that $\displaystyle \mathop{\arg \min}_{k=0,\ldots,t} \Omega^t(\d_k^t) = k_t^\star$. The desired result of AR4 then follows directly. It follows from AR2 that 
\eq
V^t(\d_{k+1}^t) - V^t(\d_k^t) \geq V^t(\d_k^t) - V^t(\d_{k-1}^t). 
\label{temp}
\en
By definition of $k_t^\star$, $C(\d_{k_t^\star+1}^t) = 2$ and $C(\d_{k_t^\star}^t) = 1$. Thus, $V^t(\d_{k_t^\star+1}^t) = V^t(\d_{k_t^\star}^t) = \Omega^{t+1}(\d_{k_{t+1}^\star}^{t+1})$. Setting $k = k_t^\star$ and then $k = k_t^\star+1$ in (\ref {temp}) we get $V^t(\d_{k_t^\star+2}^t) \geq V^t(\d_{k_t^\star+1}^t) = V^t(\d_{k_t^\star}^t) \geq V^t(\d_{k_t^\star-1}^t)$. Inductively, we conclude that $V^t(\d_{k_t^\star}^t)$ is non-increasing for $k \leq k_t^\star$ and non-decreasing for $k > k_t^\star$, as desired.

Equipped with our auxiliary results, we will show that
\begin{equation}
\begin{split}
\arg\min \{\Omega^{t+1}(\d+\v_1-\x^t), \; \Omega^{t+1}(\d+\v_2-\x^t)\} = \\
\arg\min \{\Phi(\d+\v_1-\x^t), \; \Phi(\d+\v_2-\x^t)\},
\end{split}
\label{myopicequiv}
\end{equation}
which implies that the myopic policy is optimal, because the left side of (\ref {myopicequiv}) is the decision of $\Pi_T^\star$ while the right side is the decision of the myopic policy in state $\d$ in the $t^{th}$ time-slot.

Say the switch is in state $\d$ in the $t^{th}$ time-slot and $C(\d)=2$. The states reachable from $\d$ in the $(t+1)^{st}$ time-slot are $\d_1=\d+\v_1-\x^t$ and $\d_2=\d+\v_2-\x^t$. Four cases arise, depending on whether $C(\d_1)$ and $C(\d_2)$ are 1 or 2.
\subsubsection{$C(\d_1)=2,C(\d_2)=1$}
Since $V^{T+1}(\d)=0 \; \forall \; \d$, the result is trivially true for $t=T$. Let us consider $t<T$.
It follows by definition that $V^{t+1}(\d_1) = \Omega^{t+2}(\d_1+\v_2-\x^{t+1})$ and $V^{t+1}(\d_2)=\Omega^{t+2}(\d_2+\v_1-\x^{t+1})$. However, $\d_1+\v_2-\x^{t+1}=\d_2+\v_1-\x^{t+1}=\d+\v_1+\v_2-\x^t-\x^{t+1}$. Thus, $\Omega^{t+1}(\d_1)-\Omega^{t+1}(\d_2) = \Phi(\d_1)-\Phi(\d_2)$, implying (\ref {myopicequiv}).

\subsubsection{$C(\d_1)=2,C(\d_2)=2$}
Again the result is trivially true for $t=T$. Let us consider $t<T$.
Several possibilities can arise. Since $C(\d_2)=2$, the state in the $(t+2)^{nd}$ time-slot is $\d_2+\v_2-\x^{t+1}$. The next state is determined by the optimal choice of configuration in state $\d_2+\v_2-\x^{t+1}$ in the $(t+2)^{nd}$ time-slot, and so on. In general, we can construct a {\it chain} of states which the switch visits under $\Pi_T^\star$, starting in state $\d_2$ in the $(t+1)^{st}$ time-slot. The chain terminates for one of the following two reasons:
\begin{enumerate}
\item The end of the time horizon $T$ is reached.
\item A state is reached where the optimal choice is $\v_1$.
\end{enumerate}
For all states constituting the chain except possibly the last, the optimal configuration is $\v_2$. The optimal configuration in state $\d_1$ in the $(t+1)^{st}$ time-slot is also $\v_2$. Thus, we can construct a similar chain of states originating at $\d_1$, which terminates for one of the two reasons cited above. The chain originating in state $\d_1$ comprises of the states of the type $\t_1^j = \d_1 + j\v_2 + \X^t-\X^{t+j}$ ($j=0,1,\ldots$) and the chain originating in state $\d_2$ comprises of the states of the type $\t_2^j = \d_2 + j\v_2 + \X^t - \X^{t+j}$ ($j=0,1,\ldots$).

AR3 implies that the chain originating in state $\d_1$ cannot terminate before the chain originating in state $\d_2$ due to the reason 2. AR1 implies $\Phi(\t_1^j)-\Phi(\t_2^j) \leq \Phi(\d_1)-\Phi(\d_2) \triangleq \delta_\Phi$. If both chains terminate due to reason (1), we can show $0 \leq \Omega^{t+1}(\d_1)-\Omega^{t+1}(\d_2) \leq (T-t+1)\delta_\Phi$, thereby implying (\ref {myopicequiv}). Now, suppose the chain originating in state $\d_2$ terminates in the $\tau^{th}$ time-slot ($\tau<T$) due to reason (1). We have two further sub-cases: (i) $C(\t_1^{\tau-t})=2$ and (ii) $C(\t_1^{\tau-t})=1$. For sub-case (i), we can show $0 \leq \Omega^{t+1}(\d_1)-\Omega^{t+1}(\d_2) \leq (\tau-t+1)\delta_\Phi$, thereby implying (\ref {myopicequiv}). Sub-case (ii) cannot arise because we reach a contradiction by virtue of AR4.

\subsubsection{$C(\d_1)=1,C(\d_2)=2$}
This case violates AR3 and therefore cannot arise.

\subsubsection{$C(\d_1)=1,C(\d_2)=1$}
This case leads to a contradiction similar to the one obtained in sub-case (ii) of case (2) and therefore cannot arise.

By considering a set of collectively exhaustive cases we have shown that (\ref {myopicequiv}) holds when ${\cal C}(\d)=2$. Analogous arguments can be constructed for the case ${\cal C}(\d)=1$. It follows that the optimal finite horizon policy for $N=2$ is myopic.
\end{proof}

\subsection{Proof of Lemma \ref {lemma:gammalemma}}
\label{ssec:gammalemmaproof}

\noindent\begin{proof}
The proof is by induction. We will establish monotonicity of $\gamma_{ij}$ as a function of $n_i$. The proof for monotonicity of $\gamma_{ij}$ as a function of $n_j$ follows similarly.

\subsubsection{Base Case ($t=T$)}
By definition of $\gamma_{ij}^T$ and our choice of boundary conditions,
$\gamma_{ij}^T(\n+\e_1) - \gamma_{ij}^T(\n) = \psi_i(n_i+2-\Xw_i^T)- 2\psi_i(n_i+1-\Xw_i^T)]
 + \psi_i(n_i-\Xw_i^T) \geq 0$,
where the inequality follows from the convexity of $\psi_i$.

\subsubsection{Inductive Step}
Now, assume that $\gamma_{ij}^{t+1}(\n+\e_i) \geq \gamma_{ij}^{t+1}(\n)$ for some $t<T$ and $i \neq j$. We will establish that this assumption implies $\gamma_{ij}^t(\n+\e_i) \geq \gamma_{ij}^t(\n)$. By definition,
\begin{eqnarray}
\nonumber \gamma_{ij}^t(\n+\e_i) &=& \Omega^{t+1}(\n+2\e_i) - \Omega^{t+1}(\n+\e_i+\e_j) \\
\gamma_{ij}^t(\n) &=& \Omega^{t+1}(\n+\e_i) - \Omega^{t+1}(\n+\e_j).
\label{gammadiff2}
\end{eqnarray}
Several cases arise, depending on the optimal decision in states $\n+2\e_i$, $\n+\e_i+\e_j$, $\n+\e_i$ and $\n+\e_j$ in the $(t+1)^{st}$ time-slot. Our inductive assumptions imply that the $(n_i^{t+1},n_j^{t+1})$ plane gets partitioned into $N+1$ distinct connected decision regions by the optimal policy $\Pi_T^\star$. Consequently, as $n_i^{t+1}$ increases for fixed $\{n_j^{t+1}, j \neq i\}$, $\Pi_T^\star$ {\it switches over} from ${\cal M}_i$ to ${\cal M}_k$ for some $k \neq i$. Thereafter, $\Pi_T^\star$ never switches back to ${\cal M}_i$.
This greatly restricts the number of possible cases we need to consider. We will illustrate a representative case where all four states are in the {\it interior  } of the decision region corresponding to ${\cal M}_k$. All cases in which one or more of states of interest occur at the {\it boundary} of two decision regions can be treated as a combination of the representative cases. It follows from (\ref {gammadiff2}),
\eq
\nonumber \Omega^{t+1}(\n+2\e_i) = \Omega^{t+2}(\n+2\e_i+\e_k) + \Psi(\n+2\e_i-\Xwb^{t+1}) 
\en
\eq
\nonumber \Omega^{t+1}(\n+\e_i+\e_j) = \Omega^{t+2}(\n+\e_i+\e_j+\e_k) +   \Psi(\n+\e_i+\e_j-\Xwb^{t+1}) 
\en
\eq
\nonumber \Omega^{t+1}(\n+\e_i) = \Omega^{t+2}(\n+\e_i+\e_k) + \Psi(\n+\e_i-\Xwb^t) 
\en
\eq
\nonumber \Omega^{t+1}(\n+\e_j) = \Omega^{t+2}(\n+\e_j+\e_k) + \Psi(\n+\e_j-\Xwb^t)
\en
It follows that
\eq
\nonumber \gamma_{ij}^t(\n+\e_i) = \gamma_{ij}^{t+1}(\n+\e_i+\e_k) + \Psi(\n+2\e_i-\Xwb^{t+1})  - \Psi(\n+\e_i+\e_j-\Xwb^{t+1}) 
\en
\eq
\gamma_{ij}^t(\n) = \gamma_{ij}^{t+1}(\n+\e_k) + \Psi(\n+\e_i-\Xwb^t)  - \Psi(\n+\e_j-\Xwb^t)
\en
\begin{equation}
\begin{split}
gamma_{ij}^t(\n+\e_i) &= \gamma_{ij}^{t+1}(\n+\e_i+\e_k) + \Psi(\n+2\e_i-\Xwb^{t+1}) \\  &- \Psi(\n+\e_i+\e_j-\Xwb^{t+1}) \\
\gamma_{ij}^t(\n) &= \gamma_{ij}^{t+1}(\n+\e_k) + \Psi(\n+\e_i-\Xwb^t) \\  &- \Psi(\n+\e_j-\Xwb^t).
\end{split}
\end{equation}
Also, we have from the base case,
\eq
\gamma_{ij}^T(\n+\e_i) = \Psi(\n+2\e_i-\Xwb^{t+1}) -  \Psi(\n+\e_i+\e_j-\Xwb^{t+1}) 
\en
\eq
\gamma_{ij}^T(\n) = \Psi(\n+\e_i-\Xwb^t) - \Psi(\n+\e_j-\Xwb^t)
\en
Combining, we get
\begin{equation*}
\begin{split}
\gamma_{ij}^t(\n+\e_i)-\gamma_{ij}^t(\n) &= \underbrace{\gamma_{ij}^{t+1}(\n+\e_i+\e_k)-\gamma_{ij}^{t+1}(\n+\e_k)}_{\geq 0 \text{ by our inductive assumptions}}  \\ & +  \underbrace{\gamma_{ij}^T(\n+\e_i)-\gamma_{ij}^T(\n)}_{\geq 0 \text{ by our base case}} \geq 0,
\end{split}
\end{equation*}
as desired.
\end{proof}

\subsection{Proof of Theorem \ref {theorem:msltheorem}}
\label{ssec:mslproof}

\noindent\begin{proof}
Define the time and state dependent quadratic Lyapunov function ${\cal L}^t(\d^t) \triangleq \langle \d^t, \d^t \rangle$ in state $\d^t$ in the $t^{th}$ time-slot. Given $\d^t = \d$, let $\displaystyle \v^\star = \mathop{\arg \min}_{\v \in {\cal V}} \{ \langle \d-\x^t,\v \rangle \}$. MSL($\ell$) extracts a partial configuration $\bar{\v}^\star$ from $\v^\star$ by idling all VOQs with deviation $\ell$ or more.

Given that MSL($\ell$) selects partial configuration $\bar{\v}^\star$ in the $t^{th}$ time-slot in state $\d$, the deviation vector in the $(t+1)^{st}$ time-slot is $\d^{t+1} = \d^t - \x^t + \bar{\v}^\star$. Now define the conditional one step expected drift of the Lyapunov function by
\begin{equation}
\delta_{{\cal L}}^t(\d) \triangleq \mathbb{E} \left[ {\cal L}^{t+1}(\d^{t+1}) - {\cal L}^t(\d^t) | \d^t = \d \right].
\label{drift}
\end{equation}
It follows by definition that
\begin{equation}
{\cal L}^{t+1}(\d^{t+1}) = {\cal L}^t(\d^t) + 2 \langle \d^t, \bar{\v}^\star - \x^t \rangle + \langle \bar{\v}^\star-\x^t, \bar{\v}^\star-\x^t \rangle.
\label{inter1}
\end{equation}
Conditioning on $\{\d^t=\d\}$ and taking expectation on both sides of (\ref {inter1}) we get
\begin{equation*}
\delta_{{\cal L}}^t(\d) = 2 \langle \d,\bar{\v}^\star \rangle - 2 \mathbb{E} [\langle \d, \x^t \rangle] + \langle \bar{\v}^\star, \bar{\v}^\star \rangle - 2 \mathbb{E}[ \langle \bar{\v}^\star, \x^t \rangle] + \mathbb{E} [ \langle \x^t, \x^t \rangle].
\end{equation*}
From the linearity of expectation and the definition of the inner product operator,
\begin{equation}
\begin{split}
\delta_{{\cal L}}^t(\d) &= 2 \langle \d,\bar{\v}^\star \rangle - 2\sum_{j=1}^{N^2} d_j \mathbb{E}[x_j^t] + \langle \bar{\v}^\star, \bar{\v}^\star \rangle \\ &- 2\sum_{j=1}^{N^2} \bar{v}^\star_j \mathbb{E}[x_j^t] + \sum_{j=1}^{N^2} \mathbb{E}[x_j^t x_j^t]
\end{split}
\end{equation}
Finally, invoking (\ref {iidtsp}), we have
\begin{equation}
\delta_{{\cal L}}^t(\d) = 2 \langle \d,\bar{\v}^\star \rangle -  2 \langle \d,\lambdab \rangle + \langle \bar{\v}^\star, \bar{\v}^\star \rangle - 2 \langle \lambdab, \bar{\v}^\star \rangle  + \langle \lambdab, \one \rangle.
\label{inter2}
\end{equation}
We will now bound each of the above terms individually. 

Note that $\langle \bar{\v}^\star, \bar{\v}^\star \rangle \leq N$, since a configuration vector has no more than $N$ ones. Recall that a complete configuration has exactly $N$ ones, but a partial configuration can have less than $N$ ones, if it idles some of the VOQs. 
Also, note that $\langle \lambdab, \bar{\v}^\star \rangle \geq 0$, since by definition, both the load vector and configuration vector have non-zero entries. Plugging these inequalities into (\ref {inter2}), we get
\begin{equation}
\delta_{{\cal L}}^t(\d) \leq  2 \langle \d,\bar{\v}^\star \rangle -  2 \langle \d,\lambdab \rangle + N + \langle \lambdab, \one \rangle.
\label{inter22}
\end{equation}
Consider the BV decomposition of load vector $\lambdab$, given by
\begin{equation}
\lambdab = \sum_{k=1}^{N!} \alpha_k \v_k, \quad \sum_{k=1}^{N!} \alpha_k = \alpha < 1.
\label{bvdecomp}
\end{equation}
It follows that
\begin{equation}
\langle \lambdab,\one \rangle = \sum_{k=1}^{N!} \langle \alpha_k \v_k, \one \rangle = \sum_{k=1}^{N!} \alpha_k \langle \v_k,\one \rangle = 
\sum_{k=1}^{N!} \alpha_k \cdot N =  N\alpha,
\label{inter222}
\end{equation}
where $\langle \v_k,\one \rangle = 1$ follows because $\v_k$ is a complete configuration. Next, we note that $x_i^t \in \{0,1\}$, since all entries of the target stream profiles are either 0 or 1. As a result, $\langle \d, \lambdab \rangle \geq \langle \d-\x^t, \lambdab \rangle$. Substituting for $\lambdab$ from (\ref {bvdecomp}),
\begin{equation}
\langle \d, \lambdab \rangle \geq \langle \d-\x^t, \lambdab \rangle = \sum_{k=1}^{N!} \alpha_k \langle \d-\x^t, \v_k \rangle.
\label{inter2222}
\end{equation}
The definition of the MSL service trace control policy implies
\begin{equation}
\langle \d - \x^t, \v^\star \rangle \leq \langle \d - \x^t, \v \rangle \; \forall \; \v \in {\cal V}.
\label{msldef}
\end{equation}
Combining (\ref {inter2222}) and (\ref {msldef}),
\begin{equation}
\langle \d, \lambdab \rangle \geq \sum_{k=1}^{N!} \alpha_k \langle \d-\x^t, \v^\star \rangle = \alpha \langle \d-\x^t, \v^\star \rangle.
\label{inter22222}
\end{equation}
Now, summing up both sides of (\ref {msldef}) $\forall \; \v \in {\cal V}$ and using the fact that ${\cal V} = N!$, 
\begin{equation}
N! \langle \d-\x^t,\v^\star \rangle \leq \langle \d-\x^t, \sum_{\v \in {\cal V}} \v \rangle
\end{equation}
Since each VOQ is served by exactly $(N-1)!$ complete configurations in ${\cal V}$, we have $\displaystyle \sum_{\v \in {\cal V}} \v = (N-1)! \one$, implying
\begin{equation}
\langle \d-\x^t,\v^\star \rangle \leq \frac{1}{N} \langle \d-\x^t,\one \rangle.
\label{inter222222}
\end{equation}
Since all VOQs which are idled under MSL($\ell$) have non-negative updated deviation, $\langle \d-\x^t,\bar{\v}^\star \rangle < \langle \d - \x^t,\v^\star \rangle$. Also, $\langle \x^t, \bar{\v}^\star \rangle \leq N$, since $x_i^t \in \{0,1\}$ and $\langle \bar{\v}^\star, \one \rangle \leq 1$. We now use the these observations and (\ref {inter22222}) in (\ref {inter22}) to get
\begin{eqnarray}
\nonumber \delta_{{\cal L}}^t(\d) & \leq & 2 \langle \d,\bar{\v}^\star \rangle -  2 \langle \d,\lambdab \rangle + N + \langle \lambdab, \one \rangle \\
\nonumber \; & \leq & 2 \langle \d - \x^t,\bar{\v}^\star \rangle + 2 \langle \x^t,\bar{\v}^\star \rangle  - 2 \alpha \langle \d-\x^t, \v^\star \rangle + N + N\alpha \\
\nonumber \; & \leq & 2 \langle \d - \x^t,\v^\star \rangle - 2 \alpha \langle \d-\x^t, \v^\star \rangle + 3N + N\alpha \\
\nonumber \; & \leq & 2(1-\alpha) \frac{1}{N} \langle \d-\x^t, \one \rangle + N(3+\alpha) \\
\nonumber \; & = & 2(1-\alpha) \frac{1}{N} \langle \d, \one \rangle - 2(1-\alpha) \frac{1}{N} \langle \x^t, \one \rangle + N(3+\alpha) \\
\; & \leq & 2(1-\alpha) \frac{1}{N} \langle \d, \one \rangle + N(3+\alpha)
\label{finalbound}
\end{eqnarray}
Before we can proceed further, we need the following lemma.

\begin{lemma}
If the conditional one-step drift of the quadratic Lyapunov function ${\cal L}^t(\d)$ satisfies $\delta_{{\cal L}}^t(\d^t) \leq \epsilon \langle \d^t,\one \rangle + B, \; \forall \; t>0,\d^t$ and constants $\epsilon > 0, B > 0$ (independent of state $\d^t$), then 
\begin{equation*}
\mathop{\lim \sup}_{T \rightarrow \infty} \frac{1}{T} \sum_{\tau=0}^{T-1} \sum_{i=1}^{N^2} \mathbb{E} [-d_i^\tau] \leq \frac{B}{\epsilon}.
\end{equation*}
\label{lemma:drift}
\end{lemma}

\noindent\begin{proof}
We have,
\begin{equation}
\delta_{{\cal L}}^t(\d) \leq \epsilon \langle \d, \one \rangle + B.
\label{bounddelta}
\end{equation}
Taking expectations on both sides of (\ref {bounddelta}), using the law of iterated expectations, summing up both sides for $\tau=0,\ldots,T-1$, assuming $\d^0=\zero$, and using ${\cal L} \geq 0$, we get
\begin{equation}
\frac{1}{T} \sum_{\tau=0}^{T-1} \langle -\mathbb{E}[\d^\tau], \one \rangle \leq \frac{BT}{\epsilon}.
\label{limsupineq1}
\end{equation}
Dividing both sides of (\ref {limsupineq1}) by $T$ and taking $\displaystyle \mathop{\lim \sup}_{T \rightarrow \infty}$, it follows
\begin{equation*}
\mathop{\lim \sup}_{T \rightarrow \infty} \frac{1}{T} \sum_{\tau=0}^{T-1} \sum_{i=1}^{N^2} \mathbb{E} [-d_i^\tau] \leq \frac{B}{\epsilon}.
\end{equation*}
\end{proof}
As seen from (\ref {finalbound}), the Lyapunov drift satisfies the condition of Lemma 2 with $\epsilon = 2(1-\alpha)/N > 0$ and $B = N(3+\alpha) > 0$. Since the deviation of any VOQ under the MSL($\ell$) policy is upper bounded by $\ell > 0$, we get from Lemma \ref {lemma:drift} $\forall \; j$
\begin{equation}
\mathop{\lim \sup}_{T \rightarrow \infty} \frac{1}{T} \sum_{\tau=0}^{T-1} \mathbb{E} [-d_j^t] \leq \frac{B}{\epsilon} + (N^2-1)\ell < \infty,
\label{limsupineq3}
\end{equation}
implying $\displaystyle \mathop{\lim \inf}_{t \rightarrow \infty} \mathbb{E}[d_j^t] > -\infty \; \forall \; j$, as desired.
\end{proof}

\subsection{Proof of Theorem \ref {theorem:llfsstheorem}}
\label{ssec:llfssproof}

\begin{proof}
It can be shown using Lyapunov methods that LLF guarantees finite lags to all meta-queues in the single server model of Section \ref {ssec:ss} if the inverse of the average inter-departure time targets of all meta-queues sum to less than 1. The proof of this result is very similar to the proof of Theorem \ref {theorem:msltheorem} and is therefore omitted. Bounded lags for all meta-queues in the single server model imply bounded lags for all VOQs in the switch, since the lag of a meta-queue is the maximum of the lag of its constituent VOQs (see Section \ref {ssec:single}). Under uniform i.i.d. loading, i.e., $\lambdab = \lambda \one$ for some $\lambda < 1/N$, the average inter-departure time target for every VOQ, and hence for every meta-queue in every configuration subset is $N/\lambda$. Consequently, the sum of the inverse of the average meta-queue inter-departure time targets sum to $\lambda<1$, implying the desired result.
\end{proof}

\subsection{Proof of Theorem \ref {theorem:mslsstheorem}}
\label{ssec:mslssproof}

\noindent\begin{proof}
Consider ${\cal L}^t(\d^t)$ and $\delta_{{\cal L}}^t(\d)$, as defined in the proof of Theorem \ref {theorem:msltheorem}. Assume that the switch operates in configuration subset ${\cal S}_{\v}=\{{\cal C}^k(\v)\}_{k=0}^{N-1}$. Let
\begin{equation}
k^\star = \mathop{\arg \min}_{k=0,\ldots,N-1} \langle \d-\x^t, {\cal C}^k(\v) \rangle.
\end{equation}
MSL($\ell$)-SS selects a partial configuration $\bar{\v}^\star$ which is extracted from the complete configuration ${\cal C}^{k^\star}(\v)$. Following the arguments in the proof of Theorem \ref {theorem:msltheorem}, we get from (\ref {inter22})
\begin{equation}
\delta_{{\cal L}}^t(\d) \leq  2 \langle \d,\bar{\v}^\star \rangle -  2 \langle \d,\lambdab \rangle + N + \langle \lambdab, \one \rangle.
\label{inter2ss}
\end{equation}
As always, we will bound each of these terms individually.
Since $\lambdab = \lambda \one$ for some $\displaystyle \lambda < \frac{1}{N}$ for uniform loading, it follows that $\langle \lambdab, \one \rangle = \lambda N^2$. Next, it follows from the definition of MSL-SS that
\eq
\langle \d-\x^t, \v^\star \rangle \leq \langle \d - \x^t, {\cal C}^k(\v) \rangle \; \forall \; k \in \{0,1,\ldots,N-1\}
\en
Summing both sides of the equation over $k$ we get from (\ref {subset})
\eq
N\langle \d-\x^t, \v^\star \rangle \leq \langle \d-\x^t, \sum_{k=0}^{N-1} {\cal C}^k(\v) \rangle = \langle \d-\x^t, \one \rangle \leq \langle \d, \one \rangle.
\label{inter3ss}
\en
Using $\lambdab =  \lambda \one$ once again, we get $\langle \d, \lambdab \rangle = \lambda \langle \d, \one \rangle$. Since all VOQs which are idled under the MSL($\ell$)-SS policy have non-negative updated deviation, $\langle \d-\x^t,\bar{\v}^\star \rangle < \langle \d - \x^t,\v^\star \rangle$. Also, $\langle \x^t, \bar{\v}^\star \rangle \leq N$, since $x_i^t \in \{0,1\}$ and $\langle \bar{\v}^\star, \one \rangle \leq 1$. We now use the these observations and (\ref {inter3ss} ) in (\ref {inter2ss}) to get
\begin{eqnarray}
\nonumber \delta_{{\cal L}}^t(\d) & \leq & 2 \langle \d,\bar{\v}^\star \rangle -  2 \langle \d,\lambdab \rangle + N + \langle \lambdab, \one \rangle \\
\nonumber \; & = & 2 \langle \d - \x^t,\bar{\v}^\star \rangle + 2 \langle \x^t,\bar{\v}^\star \rangle - 2 \lambda \langle \d, \one \rangle + N + N^2 \lambda \\
\nonumber \; & \leq & 2 \langle \d - \x^t,\v^\star \rangle - 2 \lambda \langle \d, \one \rangle + 3N + N^2 \lambda \\
\nonumber \; & \leq & \frac{2}{N} \langle \d, \one \rangle -2\lambda \langle \d, \one \rangle + N(3+ N\lambda) \\
	 \; & = & 2(1-N \lambda) \frac{1}{N} \langle \d, \one \rangle + N(3+ N\lambda).
\label{finalboundss}
\end{eqnarray}
Thus, we have established $\delta_{{\cal L}}^t(\d) \leq \epsilon \langle \d, \one \rangle + B$, for
$\epsilon = 2(1-N\lambda)/N$ and $B = N(3 + N\lambda) > 0$. Since the deviation of any VOQ under the MSL($\ell$)-SS policy is upper bounded by $\ell > 0$, it follows from Lemma \ref {lemma:drift} that $\forall \; j$
\begin{equation}
\mathop{\lim \sup}_{T \rightarrow \infty} \frac{1}{T} \sum_{\tau=0}^{T-1} \mathbb{E} [-d_j^t] \leq \frac{B}{\epsilon} + (N^2-1)\ell < \infty,
\end{equation}
implying $\displaystyle \mathop{\lim \inf}_{t \rightarrow \infty} \mathbb{E}[d_j^t] > -\infty \; \forall \; j$, as desired.
\end{proof}

\subsection{Proof of Theorem \ref {theorem:mslrstheorem}}
\label{ssec:mslrsproof}

\noindent\begin{proof}
Define $\delta_{{\cal L}}^t(\d,k)$ as the expected drift in the Lyapunov function ${\cal L}^t(\cdot)$ conditioned on the deviation vector $\d$ {\it and} the choice of configuration subset ${\cal S}_k$ in the $t^{th}$ time-slot. If partial configuration $\bar{\v}^\star(k)$ derived from configuration ${\cal C}^{i^\star(k)}(\v_k) \in {\cal S}_k$ is chosen in the $t^{th}$ time-slot, $\delta_{{\cal L}}^t(\d,k)$ is given by (\ref {inter2}), with $\bar{\v}^\star$ replaced by $\bar{\v}^\star(k)$. Unconditioning with respect to the choice of configuration subset ${\cal S}_k$ yields
\begin{equation}
\begin{split}
\delta_{{\cal L}}^t(\d) &=  \sum_{k=1}^{(N-1)!} \theta_k \delta_{{\cal L}}^t(\d,k) \\
&= -2 \langle \d,\lambdab \rangle - 2 \sum_{k=1}^{(N-1)!} \underbrace{\theta_k \langle {\cal C}^{i^\star(k)}(\v_k),\lambdab \rangle}_{\geq 0} \\ &+ 2\underbrace{\sum_{k=1}^{(N-1)!} \theta_k \langle \d,{\cal C}^{i^\star(k)}(\v_k) \rangle}_{W^\star} + \langle \lambdab,\one \rangle + N \\
&\leq -2\langle \d,\lambdab \rangle + 2W^\star + N(1+\alpha).
\label{interineq1}
\end{split}
\end{equation}
We will bound the terms $\langle \d, \lambdab \rangle$ and $W^\star$ to arrive at the desired result. It follows from the definition of the MSL-RS policy that
\begin{equation}
\langle \d-\x^t,{\cal C}^{i^\star(k)}(\v_k) \rangle \leq \langle \d-\x^t,{\cal C}^i(\v_k) \rangle \; \forall \; i \in \{0,1,\ldots,N-1\}.
\label{mslrsdef}
\end{equation}
Summing both sides of the equation and invoking (\ref {subset}), we get
\eq
\langle \d-\x^t,{\cal C}^{i^\star(k)}(\v_k) \rangle \leq \frac{1}{N} \langle \d,\one \rangle
\label{interrs1}
\en
Since $x_i^t \in \{0,1\}$ and a configuration vector has no more than $N$ ones, for any configuration vector $\v$, we have $\langle \v,\x^t \rangle \leq N$. Using this and the fact that $\{\theta_k\}$ is a probability distribution (implying $\displaystyle \sum_k \theta_k = 1$), we get
\eq
\sum_{k=1}^{(N-1)!} \theta_k \langle {\cal C}^{i^\star(k)}(\v_k), \x^t \rangle \leq N  \leq \sum_{k=1}^{(N-1)!} \theta_k \cdot N = N.
\label{interrs2}
\en
From (\ref {interrs1}) and (\ref {interrs2}) it follows
\eq
\langle \d,{\cal C}^{i^\star(k)}(\v_k) \rangle \leq \frac{1}{N} \langle \d, \one \rangle + \langle {\cal C}^{i^\star(k)}(\v_k), \x^t \rangle
\en
Taking the expectation on both sides of the above equation with respect to the distribution $\{\theta_k\}$ and invoking the definition of $W^\star$, we get
\eq
W^\star \leq \frac{1}{N} \langle \d,\one \rangle + N.
\label{wstar}
\en
Now, consider the BV decomposition of load vector $\lambdab$ as given by (\ref {bv}), i.e.,
\eq
\lambdab = \sum_{k=1}^{(N-1)!} \sum_{i=0}^{N-1} \zeta_{ik} {\cal C}^i(\v_k), \quad \sum_{k=1}^{(N-1)!} \sum_{i=0}^{N-1} \zeta_{ik} = \zeta < 1.
\en
Using $\langle \lambdab, \x^t \rangle \geq 0$, the definition of MSL-RS in (\ref {mslrsdef}), and the definition of $\theta_k$ in (\ref {thetadef})
\begin{eqnarray}
\nonumber \langle \d, \lambda \rangle \geq \langle \d-\x^t,\lambdab \rangle & = & \sum_{k=1}^{(N-1)!} \sum_{i=0}^{N-1} \zeta_{ik} \langle \d-\x^t, {\cal C}^i(\v_k) \rangle \\
\nonumber \; & \geq & \sum_{k=1}^{(N-1)!} \sum_{i=0}^{N-1} \zeta_{ik}  \langle \d-\x^t, {\cal C}^{i^\star(k)}(\v_k) \rangle \\
\nonumber \; &=& \sum_{k=1}^{(N-1)!} \langle \d-\x^t, {\cal C}^{i^\star(k)}(\v_k) \rangle \sum_{i=0}^{N-1} \zeta_{ik} \\
\; &=& \zeta \sum_{k=1}^{(N-1)!} \theta_k \langle \d-\x^t, {\cal C}^{i^\star(k)}(\v_k) \rangle = \zeta W^\star
\label{interss4}
\end{eqnarray}
Substituting (\ref {wstar}) and (\ref {interss4}) into (\ref {interineq1}) and noting that $\alpha = \zeta$, 
\begin{eqnarray}
\nonumber \delta_{{\cal L}}^t(\d) & \leq & -2 \zeta W^\star + 2 W^\star + N(1+\zeta)  \\
\nonumber \; &=& 2(1-\zeta)W^\star + N(1+\zeta) \\
\nonumber \; & \leq & 2(1-\zeta)\frac{1}{N} \langle \d, \one \rangle + 2N(1-\zeta) + N(1+\zeta) \\
\; & = & 2(1-\zeta)\frac{1}{N} \langle \d, \one \rangle + N(3 - \zeta)
\end{eqnarray}
We have shown that $\delta_{{\cal L}}^t(\d) \leq \epsilon \langle \d, \one \rangle + B$, for
$\epsilon = 2(1-\zeta)/N > 0$ and $B = N(3 - \zeta) > 0$. Since the deviation of any VOQ under the MSL($\ell$)-RS policy is upper bounded by $\ell > 0$, it follows from Lemma \ref {lemma:drift} that $\forall \; j$
\begin{equation}
\mathop{\lim \sup}_{T \rightarrow \infty} \frac{1}{T} \sum_{\tau=0}^{T-1} \mathbb{E} [-d_j^t] \leq \frac{B}{\epsilon} + (N^2-1)\ell < \infty,
\end{equation}
implying $\displaystyle \mathop{\lim \inf}_{t \rightarrow \infty} \mathbb{E}[d_j^t] > -\infty \; \forall \; j$, as desired.

\end{proof}

\subsection{Proof of Theorem \ref {theorem:pseltheorem}}
\label{ssec:pselproof}

\begin{proof}
Suppose MSL($\ell$) is used in the $t^{th}$ time-slot, followed by MSL-SS($\ell$) in the $(t+1)^{st},\ldots,(t+P-1)^{th}$ time-slots, and MSL$(\ell)$ again in the $(t+P)^{th}$ time-slot. We are interested in computing the $P+1$ step conditional expected drift in the Lyapunov function ${\cal L}^t(\d^t)$ of Theorem \ref {theorem:msltheorem}, given by
\begin{equation}
\begin{split}
\delta_{\cal L}^{t,P} & \triangleq \mathbb{E} [ {\cal L}^{t+P+1}(\d^{t+P+1}) - {\cal L}^t(\d^t) | \d^t=\d] \\
& = \sum_{p=0}^P \underbrace{\mathbb{E} [ {\cal L}^{t+p+1}(\d^{t+p+1}) - {\cal L}^{t+p}(\d^{t+p}) | \d^t=\d]}_{\mu_p}.
\end{split}
\end{equation}

The key idea is to bound each term $\mu_p$ individually and then obtain a bound on their sum of the form (\ref {bounddelta}). For ease of exposition, we illustrate the case $P=2$. The proof extends in straightforward fashion to $P>2$ (with more algebra).

Suppose (partial) configuration $\bar{\v}^\star_\tau$ is selected in the $\tau^{th}$ time-slot\footnote{So far, we have been suppressing the time dependence of $\bar{\v}^\star$ in all the proofs, since we were only considering one-step drifts.}. 
From Theorem \ref {theorem:msltheorem}, we have
\begin{equation}
\mu_0 \leq  2 \langle \d,\bar{\v}^\star_t \rangle - 2 \langle \d,\lambda \rangle + N(1+\alpha).
\end{equation}
From the definition of ${\cal L}$ and $\d^{t+1} = \d^t + \bar{\v}^\star_t - \x^t$, it follows
\begin{equation}
\mu_1 \leq 2 \langle \d,\bar{\v}^\star_{t+1} \rangle - 2 \langle \d,\lambdab \rangle + 3N(1+\alpha).
\end{equation}
By definition, MSL($\ell$)-SS is used in the $(t+1)^{st}$ time-slot on the subset generated by (the complete configuration corresponding to) $\bar{\v}^\star_t$. This can be used to show $\langle \d, \bar{\v}^\star_{t+1}-\bar{\v}^\star_t \rangle < 2N\alpha$. It follows that
\begin{equation}
\mu_1 \leq 2 \langle \d,\bar{\v}^\star_t \rangle - 2 \langle \d,\lambda \rangle + 3N + 5N\alpha.
\end{equation}
Next, since MSL($\ell$) is used in the $(t+2)^{nd}$ slot, it follows
\begin{equation}
\mu_2 \leq \frac{2}{N}(1-\alpha) \langle \d,\one \rangle + 7N + 11N\alpha,
\end{equation}
Finally, from Theorem \ref {theorem:msltheorem}, $\displaystyle \langle \d,\bar{\v}^\star_t \rangle - \langle \d, \lambdab \rangle \leq \frac{1}{N}(1-\alpha) \langle \d, \one \rangle + N$. Combining the above inequalities, we get
\begin{equation}
\delta_{\cal L}^{t,2}  \leq \underbrace{\frac{6}{N} (1-\alpha)}_{\epsilon} \langle \d,\one \rangle + \underbrace{15N + 17N\alpha}_{B}.
\end{equation}
The desired result now follows from arguments similar to those presented in the proof of Theorem \ref {theorem:msltheorem}. A similar bound can be established for any $P>2$, with
$\epsilon = 2(P+1)(1-\alpha)/N$ and a constant $B$ which is an increasing function of $P$.

\end{proof}

\end{document}